# The Mythical Snake which Swallows its Tail: Einstein's matter world

Galina Weinstein[1]

September 25, 2013

**Abstract:** In 1917 Einstein introduced into his field equations a cosmological term having the cosmological constant as a coefficient, in order that the theory should yield a static universe. Einstein desired to eliminate absolute space from physics according to "Mach's ideas". De Sitter objected to the "world-matter" in Einstein's world, and proposed a vacuum solution of Einstein's field equations with the cosmological constant and with no "world-matter". In 1920 the world-matter of Einstein's world was equivalent to "Mach's Ether", a carrier of the effects of inertia. De Sitter's 1917 solution predicted a spectral shift effect. In 1923 Eddington and Weyl adopted De Sitter's model and studied this effect. Einstein objected to this "cosmological problem". In 1922-1927, Friedmann and Lamaitre published dynamical universe models. Friedmann's model with cosmological constant equal to zero was the simplest general relativity universe. Einstein was willing to accept the mathematics, but not the physics of a dynamical universe. In 1929 Hubble announced the discovery that the actual universe is apparently expanding. In 1931 Einstein accepted Friedmann's model with a cosmological constant equal to zero, which he previously abhorred; he claimed that one did not need the cosmological term anymore. It was very typical to Einstein that he used to do a theoretical work and he cared about experiments and observations. This paper is a new interpretation to Einstein's cosmological considerations over the period 1917-1931.

## Introduction

In 1917 Albert Einstein decided to add an ad-hoc term to the 1916 field equations of the general theory of relativity. Einstein introduced into his general theory of relativity a cosmological term having coefficient λ, a cosmological constant, in order that the theory should yield a static model universe.[1]

At that time Einstein's general theory of relativity remained without empirical support. World War I had not yet come to the quiet. Although all scientific contact with foreign nations was disrupted by the war, Einstein maintained contact with the so-called "enemy scientists" among the allies, and with those from neutral countries, both through his link with his colleagues and best friends Hendrik Antoon Lorentz, Willem De Sitter, and Paul Ehrenfest in Leyden. Einstein was eager to check his

---

general theory of relativity as quickly as could possibly be done at that time, but he thought that his theory also required this so-called cosmological term, which neither changed the covariance of the field equations nor any other predictions of the theory.

Einstein desired to eliminate what he called the "epistemological weakness" ["erkenntnistheoretischen Schwächen"] of Newtonian mechanics, the absolute space, from physics; he invented a world, finite and spatially closed static universe, bounded in space, according to the idea of inertia having its origin in an interaction between the mass under consideration and all of the other masses in the universe, which he called "Mach's ideas" (obviously not Ernst Mach's ideas as has been generally recognized and as Mach himself pronounced them). This would be later called by Einstein "Mach's principle" (more precisely Mach-Einstein principle).

Physically then the cosmological constant $\lambda > 0$ implies the existence of cosmic repulsion, Einstein's static universe being one in which this repulsion everywhere counterbalances gravitational attraction.[2]

Einstein envisaged his cosmological model because the $g_{\mu\nu}$ were all determined by the field equations, of which the stress-energy tensor depends on matter. Thus matter also appears as the source of the $g_{\mu\nu}$, i.e. of inertia. The problem was then (and De Sitter discussed it with Einstein): can we say that the whole of the $g_{\mu\nu}$ is derived from these sources? The field equations determine the $g_{\mu\nu}$ apart from boundary conditions, which can be mathematically defined by stating the values of $g_{\mu\nu}$ at infinity. Einstein hesitated and finally decided to abolish the boundary conditions and invoked a finite and spatially closed universe, bounded in space.

De Sitter objected to this solution because of the "world-matter" density in Einstein's universe, which was related with the cosmological constant; ordinary matter density was nebulae (galaxies in later terminology), stars, and so on. De Sitter proposed a vacuum solution of Einstein's field equations with the cosmological constant and with no "world-matter". De Sitter's world was thus empty. In De Sitter's vacuum solution of the modified field equations the cosmological term did not depend on any world-matter that was of course not present in his universe. De Sitter proposed a solution by assuming that world-matter density equals zero. In this sense his solution was an empty universe.

De Sitter's world was spherical in its space dimensions, but open towards plus and minus infinity in its time dimension (if real time was used), like an hyperboloid. De Sitter become aware of the experimental work by Vesto Melvin Slipher possessing the radial velocities of twenty-five spiral nebulae; yet in 1917 De Sitter knew of only three of them. He then proposed a redshift effect in his hyperboloid world.

In order better to compare his own model with Einstein's solution, De Sitter wrote his solution in a static form. He compared between both models, which he called A (Einstein's) and B (De Sitter's) by using spherical polar coordinates.

De Sitter explained that, in solutions A and B three-dimensional space has constant positive curvature, and the system B is the four-dimensional analogy of the three-dimensional space of the system A. In B the time is entirely relative, and completely equivalent to the other three coordinates, but Einstein's whole world A is filled homogenously with matter, and at infinity t' = t, and thus system A introduces "a quasi-absolute time". In addition, the system A only satisfies the postulate of relativity if the latter is applied to three-dimensional space. The world-matter thus takes the place of the absolute space in Newton's theory.

Einstein was hard to give up Mach's principle; he tried instead to demonstrate that De Sitter's solution contained a singularity, and thus argued that De Sitter's model was actually not matter-free. According to general relativity, the closer clocks are to a material source, the more slowly they run; Einstein thus reasoned that, clocks slowed down as they approached the equator of De Sitter's solution in the static form, and all matter (world-mater) of De Sitter's world was concentrated there. Einstein concluded that De Sitter's solution contained an intrinsic singularity indicating there is hidden matter in the equator.

Felix Klein demonstrated to Einstein that the equator in the static form of the De Sitter solution is an artifact of the static form: the equator can never be reached, because the coordinate system in which the De Sitter world is static covers only part of the entire De Sitter space-time. The singularity at the equator can be transformed away and does not indicate the presence of matter.

 At the end Einstein became convinced that the De Sitter solution was indeed a solution to his modified field equations, but he still believed that it was not a physical possible world, because he held that any acceptable cosmological model would have to be static.

In 1919 Einstein realized that he found a marvelous idea which was significant *for his then plan to getting rid of the horrifying ad-hoc character of the cosmological term*. The cosmological constant $\lambda$ arises as an integration constant in the solution of certain equations, which the matter tensor is naturally required to satisfy in general relativity.[3]

In 1920, however, Einstein introduced "Mach's Ether". The world-matter of Einstein's world was equivalent to an ether, a Machian substance that was needed as a carrier of the effects of inertia. The world-matter was certainly essential, because as he said, the modern physicist does not believe that he may accept action at a distance, he comes back once more, if he follows Mach, to the ether, which has to serve as medium for the effects of inertia.

In April-May 1921 Einstein joined Chaim Weizmann's tour of the United States to gain support among American Jewry for the Zionist cause. His role was to raise funds for the establishment of Hebrew University of Jerusalem. Princeton University has arranged five lectures on the theory of relativity on the afternoon from May 9 to 13, 1921, the subject of these lectures, which were delivered in German, were special relativity, general relativity, "Generalities on the Theory of Relativity", and "Cosmological Speculations". The last lecture given the afternoon of May 13 dealt with general relativity and cosmology. Einstein talked in McCosh 50 hall at 4:15 in the afternoon. A *New York Times* Reporter described Einstein's universe, as spherical of finite extent, but infinite because of its curved nature and wrote that Einstein conceives the universe as being bent back upon itself much as the mythical snake which swallows its tail.

In De Sitter's 1917 solution gravitation is negligible compared with cosmic repulsion, thus predicting recession of galaxies. In 1922-1923 Arthur Stanley Eddington and independently Hermann Klaus Hugo Weyl indicated that Einstein's cosmological term was related to a non-static element in De Sitter's world. They found support for De Sitter's model in Slipher's observed radial velocities of spiral nebulae.[4] On the basis of Slipher's experimental observations, Weyl thus rejected Einstein's model and chose De Sitter's; he obtained a relation between redshift and distance in De Sitter's universe, which was established using "Weyl's principle".

In 1923 Einstein sent a postcard to Weyl in which he wrote: "Regarding the cosmological problem I do not agree". He then said that according to De Sitter two points are in a motion of recession, and if there is no quasi-static world then away with the cosmological term.[5]

The "cosmological problem" that Einstein did not agree with was probably *the De Sitter spectral shift effect.*

In 1917 Einstein claimed that small velocities of the stars (with respect to the velocity of light), $v << c$, allow us to assume *a static universe*. In Einstein's static universe no redshift of the spectral lines occurs unless the source moves with a velocity relative to the receiver. This result is consistent with the Doppler effect. If then a star moves through an otherwise static background of matter then it produces a redshift. Silpher's redshift data still suggested that $v << c$, allowing a quasi-static universe. In the De Sitter world, with the cosmological constant, we see a contribution to the redshift even if the emitting star does not move relative to the observer. It then looks as if the stars are accelerating away from the observer and are in a motion of recession. This result is the De Sitter spectral shift effect. In De Sitter's universe if all stars were supposed not to exist, with the exception of one single star placed at a huge distance from the observer, then the geometry of the world would have had effects on the signal sent from the source to the receiver.[6]

De Sitter's model was not without its difficulties; for instance, it had the strange De Sitter spectral shift effect. For Einstein its major weakness, to the point of apparent fatality, was simply that it violated his 1918 Mach's principle. If thus De Sitter's world was found to be non-static, then Einstein thought that there was no point in keeping the cosmological constant. After all Einstein introduced into his field equations the cosmological term having the cosmological constant as a coefficient, in order that the theory should yield a *static universe*.

In 1922, Aleksandr Friedmann and in 1927 Georges Édouard (Abbé) Lamaître published independently two dynamical universe models. Friedmann discovered interesting non-static models with a cosmological constant equal to zero or not equal to zero. This was a prediction of an expanding or a contracting universe of which Einstein's and de Sitter worlds were special cases. Friedmann's model with a cosmological constant equal to zero was the simplest general relativity universe. In 1931 Einstein adopted this view and dropped the cosmological term publicly.

But in 1922 Einstein thought he had found a mistake in Friedmann's results, which when corrected Friedmann's solution would give Einstein's static model. Friedmann sent Einstein his calculations, and Einstein became convinced by Friedmann's letter communicated to him that, Friedmann's results were mathematically correct, but he thought that Friedmann's suggestion was not a physical possible model. At about the year 1922, for good or ill, Einstein was unwilling to abandon his cosmological model, and was little impressed by Friedmann's mathematical models. We can almost hear Einstein thinking aloud that Friedmann's work is just extra mathematical complication.

In 1925 Lemaître suggested a modification of De Sitter's world which included a non-static character and dependence upon distance of the redshift of line spectra caused by the Doppler effect, which he quickly gave up by 1927. In 1927 he independently published quite the same model of the expanding universe as Friedmann's; but Lemaître's model was "more astronomical" than Friedmann's mathematical model.

When Lemaître published his paper in 1927 he was unaware of the little known papers of 1922 and 1924 by Friedmann. Lemaître became aware of Friedmann's work when he had a face-to-face encounter with Einstein in the 1927 Solvay Conference in Brussels. Einstein's response to Lemaître's work indicated the same unwillingness to change his position that characterized his response to Friedmann's work. Einstein was willing to accept the mathematics, but not the physics of a dynamical universe.

In 1929 Edwin Hubble announced the discovery that the actual universe is apparently expanding. In the years beyond 1930, the tide turned in favor of dynamical models of the universe. The discovery was hailed as fulfilling the prediction of general relativity.[7]

In 1931 Einstein became aware of this revolution during a visit to Caltech in Pasadena. Upon his return to Berlin the new experimental and theoretical findings have led Einstein to drop his old suggestions in favor of new ones, the dynamical universe. Einstein returned to the unmodified field equations of general relativity; he accepted the dynamical case, Friedmann's model with a cosmological constant equal to zero, which he had previously abhorred, and he claimed that one did not need the cosmological term for this case. He found that models of the expanding universe could be achieved without any mention of the cosmological constant. *It was very typical to Einstein that he used to do a theoretical work and he cared about experiments and observations.*

Einstein published a short paper in April 1931, "On the Cosmological problem of General Relativity"; in this paper he studied Friedmann's non-static solution of the field equations of the general theory of relativity, of which the line element corresponded to a cosmological constant equal to zero. In an August 1931 paper De Sitter adopted this line element with a cosmological constant equal to zero. A few months later, on January 1932, when both De Sitter and Einstein were visiting Mount Wilson Observatory, they wrote a joint paper in which they presented the Einstein-De Sitter universe following Einstein's lead without the cosmological term. On the basis of the model that was studied by Einstein in 1931, without the introduction of both a cosmological constant and a curvature, Einstein and De Sitter obtained the dependence of the coefficient of expansion on the measured redshifts.

In 1931-1932 thus new experimental and theoretical findings led Einstein finally to drop his old suggestions in favor of new ones. We usually characterize Einstein's renouncement of the cosmological constant and coming up with new ideas as Einstein's mistake. Perhaps we rather say that Einstein's old and new ideas link up with the same good old Mach's principle; later models of Einstein are either compatible or incompatible with Einstein's understanding of this principle.

On the other side of the Atlantic, in 1930 Arthur Stanley Eddington inferred that the universe had emerged from the Einstein static state. He said that instead of having to choose between Einstein's and de Sitter's worlds, the universe started as an Einstein world, being unstable it began to expand, and it is now progressing towards de Sitter's form as an ultimate limit. His ideas had little influence upon the subsequent history of cosmology.

In 1933 Lemaître proposed the first (modern) history of the universe. Following Einstein's lead, any use of the cosmological constant was generally out of favor, and therefore Lemaître's ideas were given less attention than they deserved. William McCrea described Lemaître's idea: Taking the cosmological constant > 0, Lemaître found an expression factor satisfying the Friedmann-Lemaître equations for the relativistic homogeneous isotropic expanding cosmological model. The Lemaître singularity at t = 0, followed by rapid expansion, this being decelerated by self-gravitation leading to near-stagnation in the vicinity of the Einstein static state, independent of t, if the value of the cosmological constant is suitably chosen, until the onset of accelerated expansion under cosmic repulsion. Lemaître pictured the very early universe as a primeval atom, cosmic atomic nucleus, with the big bang as its spontaneous radioactive decay. Thus the very early universe would have been dominated by high-energy particles producing a homogenous early universe. Cosmic rays were inferred to be the most energetic relict particles from the decay, so that they constituted background radiation for the model.[8]

The elder Einstein could not remember how far Mach's writings have influenced his work in the same way as could the young Einstein who was inspired by Mach's ideas when creating the general theory of relativity. Indeed the elder Einstein often wrote that the influence of David Hume was greater on him. Finally, a year before his death Einstein silently dropped Mach's principle in itself.

This paper concentrates on Einstein's "cosmological" journey through a rather "bumpy road" over the period 1917-1931.

### Einstein's First Trials and De Sitter's 1916 Objections

In **February 1916** word reached Einstein that Mach died and he wrote an obituary for Mach.[9] A month later, in **March 1916** Einstein sent to the editor of the *Annalen der Physik*, Wilhelm Wien, a review article on the general theory of relativity. Einstein completed the paper on **20/3/1916**. The paper was published two months later, in **May 1916**.[10]

By the time this paper was published, Einstein had already produced the main scaffold of his new cosmological ideas, of which he described very sketchily to Michele Besso in a letter from **May 14, 1916**. Einstein told his best friend that, "in gravity I was now trying to determine the boundary conditions at infinity". He thought it would be "interesting to think about the possibility of a *finite* world, i.e. spatially finite world, in which of course, all inertia is really relative".[11]

Over the period 1915-1916 Einstein's German publications concerning general relativity did not reach countries at war with Germany, but they continued to reach neutral countries; in particular scientists in Leiden and of course Einstein's close friends, Hendrik Antoon Lorentz, Paul Ehrenfest, and De Sitter, were well-informed.

De Sitter must have been among the first to assimilate Einstein's work and to see its significance. He sent a comprehensive explanation of it in the form of three substantial papers entitled "On Einstein's theory of gravitation and its astronomical consequences" totaling almost 90 pages for publication in *Monthly Notices of the Royal Astronomical Society*. Arthur Eddington was the Secretary who received these papers and saw to their publication.[12] De Sitter began his first paper, communicated to the Journal in **August 1916**, with Einstein's "'Allgemeine Relativitätstheorie' of 1915, by which, moreover, gravitation is also incorporated in the union".[13]

In **June 1916** De Sitter heard from Einstein himself about his cosmological ideas. He stringed a few words together about Einstein's solution of the problem of boundary conditions in the second paper of **(September) October 1916**.[14]

Newton's integral law of gravitation was replaced by Poisson's equation, the concept of a field of gravitation, of which the material bodies were the sources, was introduced. At very large distances from any material body the potential is zero: the whole of the numerical value of the potential at any point can thus be derived from the material sources. Expressed in the language of Einstein's general theory of relativity, we would say, at infinity the $g_{\mu\nu}$ are the Minkowski flat metric. These are the natural values. And Einstein explains that it is following strictly Mach's ideas: "It seems, therefore, that such a degeneration of the coefficients $g_{\mu\nu}$ at infinity is required by the postulate of relativity of all inertia".[15]

In Einstein's theory all the $g_{ij}$ differ from the values of the Minkowski's special relativity metric, and they are all determined by field equations, of which the right-hand members $kT_{ij}$ depend on matter. Thus matter here also appears as the source of the $g_{ij}$, i.e. of inertia. But not the whole of the $g_{ij}$ is derived from these sources. The differential equations determine the $g_{ij}$ apart from constants of integration, or boundary conditions, which can be mathematically defined by stating the values of $g_{ij}$ at infinity. And De Sitter explains that a particular consequence of Einstein's above assumption is its violation of the principle of relativity: [16]

"Evidently we could only say that the whole of the $g_{ij}$ is of material origin if these values at infinity were *the same for all systems of co-ordinates*. It is not necessary to insist on the values [of the flat Minkowski metric]. We can prescribe any other set of degenerated values which the $g_{ij}$ must have at infinity, only they must be the same for all systems of reference, i.e. they must be invariant for all transformations, or at least for a so comprehensive group of transformations that restriction to this group does not mean giving up the principle of relativity. The values of the [Minkowski flat metric] are certainly not invariant".

Einstein understood that De Sitter had some grounds for his criticism, and he modified his suggestion. In **September 1916** Einstein visited Leiden. While repeating what from the relativity principle perspective was the same fault as before (retaining a so-called absolute space/time element), Einstein in conversations with De Sitter: "has however, pointed out a set of degenerated $g_{ij}$ which are actually invariant for all transformations in which, at infinity, $x_4$ is a pure function of $x_4'$. They are [a metric in which the time t' at infinity is a function of t, but not of the special coordinates $x_1$, $x_2$, $x_3$]". Einstein makes the hypothesis that at infinity the $g_{ij}$ actually have these values. The meaning of this degenerated set of $g_{ij}$ is that at infinity the four-dimensional space-time is dissolved into a three dimensional space and a one-dimensional time.

De Sitter, of course, could not accept this attempt to modify the previous suggestion, because "at infinity we would thus have an absolute time, but no absolute space". According to Einstein's hypothesis "there must exist, at still larger distances, certain unknown masses which are the sources of the values" of the Minkowski metric, "i.e. of all inertia. These hypothetical masses would form the boundary of the universe, which would thus be necessarily finite and limited. Outside them there would be *nothing* but the field of $g_{ij}$ gradually degenerating to the values" of Einstein's metric in which the time t' at infinity is a function of t, but not of the special coordinates $x_1$, $x_2$, $x_3$.

Einstein was thinking in terms of Mach's ideas, but De Sitter held that a consequence of the field equations was that they admit another interpretation: "We must insist on the impossibility that any of the known fixed stars or nebulae can form part of these hypothetical masses".

Einstein responded on **February 2, 1917**. He understood that De Sitter had some grounds for his criticism, and he modified his presentation: "presently I am writing a paper on the boundary conditions in gravitation theory".

He informed De Sitter that he completely abandoned his previous idea which De Sitter rightly disputed, and he was curious to see what De Sitter would say about his new outlandish idea he had now set his sights on. [17]

In **March 1917**, De Sitter prepared the final paper of the trilogy, "On Einstein's theory of gravitation and its astronomical consequences" as a response to Einstein's above outlandish conception, and in **July 1917**, he communicated it to the *Monthly Notices of the Royal Astronomical Society*.[18]

De Sitter raised the following objections:[19]

"If all matter were destroyed, with the exception of one material particle, then would this particle have inertia or not? The school of Mach requires the answer *No*. If, However, by 'all matter' is meant all matter known to us, stars, nebulae, clusters, etc, then the observations very decidedly give the answer *Yes*. The followers of Mach are therefore compelled to assume the existence of still more matter. This matter, however, fulfils no other purpose than to enable us to suppose it not to exist, and to assert that in that case there would be no inertia. This point of view, which denies the logical possibility of the existence of a world without matter, I call the *material postulate of relativity if inertia*. The hypothetical matter introduced in accordance with it I call *world-matter*. Einstein originally supposed that the desired effect could be brought about by very large masses at very large distances. He has, however, now convinced himself that this is not possible. In the solution which he now proposes, the world-matter is not accumulated at the boundary of the universe, but distributed over the whole world, which is finite, though unlimited. Its density (in natural measure) is constant, when sufficiently large unites of space are used to measure it. Locally its distribution may be very unhomogeneous. In fact, there is no essential difference between the nature of ordinary gravitating matter and the world-matter. Ordinary matter, the sun, stars, etc., are only condensed world-matter, and it is possible, though not necessary, to assume all world-matter to be so condensed. In this theory 'inertia' is produced by the whole of the world-matter, and 'gravitation' by its local deviations from homogeneity.

In Einstein's new solution the three dimensional world is not infinite, but spherical. Thus no boundary conditions at infinity are required".

And De Sitter advanced the same criticism in the other paper: [20]

"Einstein has not succeeded in finding such a set of boundary values [which are the same for all systems of reference] and therefore makes the hypothesis that the universe is not infinite, but spherical: then no boundary conditions are needed, and the difficulty disappears".

In view of the foregoing objection, on **February 2, 1917** Einstein wrote De Sitter: "I have completely abandoned my views, rightfully contested by you, on the degeneration of the $g_{\mu\nu}$. I am curious to hear what you will have to say about the somewhat crazy idea I am considering now".

**Einstein's Finite, Spatially Closed Universe and the Cosmological Constant 1917**

While ruminating and brooding on his new (cosmological) idea, Einstein felt as if he was a little mad. He wrote Paul Ehrenfest on **February 4, 1917** about this feeling: "I have again perpetrated something relating to the theory of gravitation that might endanger me of being committed to a madhouse".[21]

Four days later, on **February 8, 1917** Einstein's paper "Cosmological Considerations in the General Theory of Relativity" was published.[22]

The assumption upon which Einstein's new solution rests is that, no boundary conditions at spatial infinity are required. Einstein indeed ventured to suggest a three dimensional world, not infinite, but bounded and spherical.

Einstein started the discussion of cosmological model with the Newtonian world. Newtonian classical universe just happened to be plugged with exactly the crucial problem and needed just the correction at the right time. As already mentioned by Hugo Hans Ritter von Seeliger and Carl Neumann for Newtonian gravitational theory (known as the "Neumann-Seeliger paradox"), a static infinite universe filled with matter needed a repulsive force that varied with the distance.[23] Thanks to this problem in Newtonian theory, Einstein could save the relativistic ad-hoc constant that he was going to introduce. Apparently, the relativistic cosmological constant could be analogous to its Newtonian counterpart: it represented a repulsive force that counterbalanced gravitational attraction.

Einstein then started by asking: how far could the Poisson equation in Newtonian theory

$$\nabla^2\phi = 4\pi K\rho$$

be preserved for an infinite universe with uniformly matter distribution? [24]

Einstein expressed his opinion about the problem: We must, supplement Poisson's equation and the Newtonian equations of motion of material points by boundary conditions at spatial infinity: Newtonian theory permits a universe with a central distribution of matter and a boundary region of empty space. This condition leads to the view that the density of matter ρ becomes zero at infinity.

Newton's entire universe must be finite, but there can be an infinite amount of mass in this universe; an infinite spherical surface universe. Einstein cancelled the possibility of empty space in Newtonian universe, because of Mach's ideas. And by close analogy he did the same in relativistic universe.

Of course Einstein noticed that Newtonian finite universe entailed serious consequences; it had a tendency for instability. It was puzzling that under the conditions of static (no global motion) world a heavenly body with kinetic energy was able to reach spatial infinity by overcoming the Newtonian forces of attraction. Einstein based himself on statistical mechanics, and reasoned that if the total energy of the stellar system, transferred to a single star, as a result of random motions, was great enough to send that star on its journey to infinity, then it might never return. Thus in this world heavenly bodies would attract each other to form clusters, and mass distribution would vary accordingly and not stay homogenous for a long time; there are attractive forces (gravity) involved and no repulsive forces counter-balancing the gravitational forces. [25]

Einstein did not express doubts about the Newtonian world being static; he desired a static Newtonian universe and this paved the way for his own cosmological universe. Einstein's solution to this problem was modification of Poisson' s equation by adding a term:

$$\nabla^2 \phi \ - \lambda \phi = 4\pi G\rho$$

Here λ denotes a universal constant. The term $\lambda\phi$ depends on the uniform (constant) density of the distribution of mass $\rho_0$:

$$\phi = -\frac{4\pi G}{\lambda}\rho_0$$

and it acts as a repulsive force. [26]

If $\rho = \rho_0$ we obtain:

$$\nabla^2 \phi \ = 0$$

This is the Laplace equation. Hence ρ is constant and matter is uniformly and homogenously distributed in the universe.

Insofar as λ denoted a cosmic constant in Newtonian world, Einstein's explanation was plausible. He was now set about to proceed to his own theory and modify the field equations in much the same way as he had done with the Poisson equation. The field equations should be treated effectively the same as Poisson's equation.

Firstly, Einstein investigated (as already stated, about summer 1916) centrally symmetrical, static gravitational fields, degenerating at infinity. He explained that suppose we consider boundary conditions according to which space-time is flat, In this case, asymptotically Riemann's tensor vanishes in spatial infinity. We therefore obtain a flat Minkowski space-time solution to the field equations, quasi-Euclidean space. This solution violates Mach's ideas (Mach's principle) of the relativity of inertia, because inertia would not be conditioned by matter: "In a consistent theory of relativity there can be no inertia *relatively to 'space'*, but only an inertia of masses *relatively to one another*. If, therefore, I remove a mass to a sufficient distance from all other masses in the universe, its inertia must fall to zero".[27] Inertia of mass points should depend on $g_{\mu\nu}$, and these must be determined by the distribution of matter in the universe.

Secondly, Einstein explained that, the same objections must be raised against general relativistic infinite static universe which are raised in respect to the Newtonian Universe. Einstein, not being able to formulate boundary conditions at spatial infinity, gave up completely the boundary conditions at spatial infinity. Instead he chose a finite, closed universe with respect to its spatial dimensions. So at any rate the first step was modifying the field equations.

For several reasons Einstein thought that we must willingly accept the modifications to which he brings us: The small velocities of the stars allowed Einstein to assume that whenever there are fixed stars, the gravitational potential is not bigger than that on earth. Hence, he assumed a system of reference with respect to which matter (fixed stars) is permanently at rest, a static universe; $\rho$ the mean density of matter in Einstein's universe is a scalar and is function of the space coordinates. The time coordinate $x_4$ was independent for all magnitudes, so that for the required solution for all $x_4$, $g_{44} = 1$. Masses generating the field are uniformly distributed and the curvature of the space is constant. The three-dimensional spatially finite world of the $x_1$, $x_2$, $x_3$, with constant $x_4$, was a spherical homogenous and isotropic space.

Einstein now showed that, if the field equations were modified by introducing a (cosmological) constant, the spatially closed universe with uniform mass distribution, otherwise unstable, could now be simply stabilized with the help of this cosmological term.

The unmodified field equations of gravitation are: [28]

(1) $G_{\mu\nu} = -\kappa \left( T_{\mu\nu} - \frac{1}{2} g_{\mu\nu} T \right)$

This field equation does not yield a static solution. Therefore, Einstein added on the left-hand side of this field equation the tensor $g_{\mu\nu}$ multiplied by the universal infamous cosmological constant $\lambda$ ($\lambda$ being sufficiently small):[29]

$$(2) \quad G_{\mu\nu} - \lambda g_{\mu\nu} = -\kappa \left( T_{\mu\nu} - \frac{1}{2} g_{\mu\nu} T \right)$$

This did not destroy the general covariance of the field equations and the equations were compatible with the relativity principle. The laws of conservation of momentum and energy were also satisfied.

The value of the cosmological constant is: [30]

$$(3) \quad \lambda = \frac{1}{R^2} = \frac{\kappa \rho}{2} = \frac{4\pi G \rho}{c^4}, \kappa = \frac{8\pi G}{c^4}$$

with $\kappa$ "gravitational constant" in field equations and R the radius of Einstein's three-dimensional spatial universe. Hence, the cosmological constant was related in a simple way to the mean density of matter $\rho$ of matter in the Universe.

This model fully agreed with the relativity of inertia, but Einstein understood that a consequence of his world was that it admitted an interpretation effectively in terms of absolute time. On **February 14, 1917**, he wrote Paul Ehrenfest:[31]

"I am sending you my new paper. My solution may appear adventurous to you, but for the moment it seems to me to be the most natural one. From the measured stellar densities, a universe radius of the order of magnitude 107 light years results, thus unfortunately being very large against the distances of observable stars. The odd thing is that now a quasi-absolute time and a preferred coordinate system do reappear in the end, while fully complying with all the requirements of relativity. Please show the paper also to Lorentz and de Sitter".

Moreover, things were going to be different when models other than Einstein's were to be invented; and when it later came to experimental evidence in favor of Einstein's beautiful static castle, no evidence was available at that time and Einstein could not say whether his static universe corresponded to reality. Remember that this was about ten years before the major experimental astronomical discoveries of Hubble, and Einstein as usual had no means other than his free creation of the mind. On **March 12, 1917** Einstein wrote De Sitter: "From the standpoint of astronomy, I have, of course, built there a spacious castle in the air".[32]

### De Sitter's "Empty" World

On **March 31, 1917** De Sitter submitted his paper "On the relativity of inertia. Remarks concerning Einstein's latest hypothesis" in which he put forward a different suggestion depending upon non-Machian ideas. De Sitter wished to show that the Machian view, enrooted in Einstein theory, was in fact not crucial for a consistent cosmology. Earlier, on **March 26, 1917**, he communicated his paper, "On the relativity of inertia. Remarks concerning Einstein's latest hypothesis", to the Proceedings of the Academy of Amsterdam.[33]

De Sitter attempted to show that Einstein makes the hypothesis that his world is spherical, but from the point of view of the theory of relativity it appears at first sight to be incorrect to say: the world is spherical, for it can by a transformation be represented in a Euclidean space. This is a perfectly legitimate transformation, which leaves the line element unaltered. But even this shows that also in the Euclidean system of coordinates Einstein's world remains finite and spherical. All the $g_{\mu\nu}$ for this spherical world are transformed to a set of values which at infinity degenerate to special values which are zero, but temporal value which is 1. Einstein then assumes that his world is a finite three-dimensional universe rather than a four-dimensional one and that is the reason for why $g_{44} = 1$ (the so-called "absolute time" element in Einstein's world). But he has no other choice, because he assumes according to Mach's ideas that his three-dimensional space is filled with matter. The $g_{\mu\nu}$ of Einstein's spherical world and their transformed values in the Euclidean system of reference do not satisfy Einstein's unmodified or original (1916) field equations, for these values cannot be the same in all systems of reference. That is the reason for why Einstein is obliged to add the cosmological term. However, De Sitter is not obliged to Mach's ideas and he thus extends Einstein's hypothesis to the four-dimensional space-time, and then finds that the $g_{\mu\nu}$ all degenerate at infinity to zeros, including the $g_{44}$.[34]

De Sitter suggests a "complete relativity of the time" world solution of Einstein field equations with the cosmological constant, a counterexample to what De Sitter calls Einstein's "world-matter". When at infinity all $g_{\mu\nu}$ are zero "*I have called this the mathematical postulate of relativity of inertia*".[35]

De Sitter asserts:[36]

"Einstein's solution of the [field] equations implies the existence of a 'world-matter' which fills the whole universe, as has already been mentioned. It is, however, also possible to satisfy the equations without this hypothetical world-matter. Then, of course, the 'material postulate of relativity of inertia' is not satisfied, but the 'mathematical postulate', which makes no mention of matter, but only requires the $g_{\mu\nu}$ to be zero at infinity, is satisfied. This is brought about by the introduction of the term with $\lambda$, and not by the world-matter, which, from this point of view, is not essential".

When at infinity all $g_{\mu\nu}$ are so defined, the departure from Einstein's "Mach's ideas" is fulfilled. And De Sitter concluded:[37] "we find the remarkable result, that now no 'world-matter' is required".

Rather than assuming that only space was finite, De Sitter assumed that space-time was finite. In De Sitter's vacuum solution the cosmological constant was unrelated with the matter density, "world matter" $\rho$. In Einstein's world, world-matter density was related with the cosmological constant. Thus De Sitter proposed a solution by assuming that world-matter density equals zero. In this sense De Sitter's solution was an empty universe.

De Sitter's world allowed the following possibility: [38] "To the question: If all matter is supposed not to exist, with the exception of one material point, which is to be used as a test-body, has then this test-body inertia or not? The school of Mach [Einstein's model] requires the answer *No*. Our experience [De Sitter's model] however very decidedly give the answer *Yes*, if by 'all matter' is meant all ordinary physical matter: stars, nebulae, clusters, etc".

De Sitter agreed with Einstein[39] that his "world-matter" three-dimensional space universe was nothing but a reformulation of absolute space ("preferred coordinate system").[40] Paradoxically, Einstein's intention in general relativity was to drop absolute space and time and chose the relativity principle, but in 1917 he preferred Mach's principle and almost came back again to the same starting point of absolute space with his cosmological model. Hence, De Sitter suggested:[41] "We can also abandon the postulate of Mach, and replace it by the postulate that at infinity the $g_{\mu\nu}$ […] of three-dimensional space, shall be zero, or at least invariant for all transformations".

And De Sitter added: "The introduction of this constant [cosmological constant] can only be avoided by abandoning the postulate of the relativity of inertia all together".[42] On **March 20, 1917** he wrote Einstein that, he preferred the four-dimensional world, and even more than that, "the original theory, without the undeterminable λ, which is just philosophically and not physically desirable".[43]

De Sitter's world was obtained with Einstein's field equations and the cosmological term. But De Sitter's world was empty, i.e., no "matter-world". Therefore the cosmological term did not depend on any matter-world that was of course not present in De Sitter's universe.

This was surely true, but for his part Einstein did not agree with this claim. According to Einstein De Sitter had no grounds for his suggestion and for his standpoint; Einstein was not even willing slightly to modify the suggestion given in his February paper. Einstein thought that his cosmological model was proof against De Sitter's objection, because of Mach's principle.

On **March 24, 1917** Einstein replied to De Sitter's letter and explained: "The $g^{\mu\nu}$-field should rather *be determined by the matter, and not be able to exist without it*. This is the essence of what I understand by the demand for the relativity of inertia. […] As long as this demand was not fulfilled, for me the goal of general relativity was not yet completely achieved. This was first achieved through the introduction of the λ term."[44]

### Einstein Objects to the static form of De Sitter's solution

In 1917, opinions could differ as to Mach's ideas, but in **March 1917** De Sitter compared between the two systems, Einstein's and his own model.

The *spatial* geometry of Einstein's world is a three-dimensional hypersphere of radius R embedded in a four-dimensional Euclidean space with positive constant curvature; the hypersphere is defined by:

$$R^2 - x^2 - y^2 - z^2 - t^2 = 0.$$

De Sitter attempted to: "first take the system of reference used by Einstein. In case A [Einstein's model] we take: $x_4 = ct$, in B [De Sitter's model] I take, for the sake of symmetry, $x_4 = ict$. In both cases R is the radius of curvature of the hypersphere".[45]

The *space-time* geometry of De Sitter's world is then a four-dimensional hypersphere embedded in a five-dimensional Euclidean space. "The system B is the four-dimensional analogy of the three-dimensional space of the system A".[46]

Using real time coordinate, the world is a four-dimensional hyperboloid in a 4 + 1 dimensional Minkowski spacetime with constant negative curvature. It is spherical in its four spatial dimensions, but open towards plus and minus infinity in its temporal dimension, like an hyperboloid. In the limit of zero curvature, when the radius R tends to infinity, we obtain Minkowski's space-time.[47] De Sitter explained:[48]

"We have

$$ds^2 = -dx^2 - dy^2 - dz^2 + dt^2 - du^2,$$

and

$$(a)\ R^2 - x^2 - y^2 - z^2 + t^2 - u^2 = 0.$$

The latter equation represents an hyperboloid (one bladed) in the five-dimensional space (x, y, z, t, u). The projection of a point x, y, z, t, u of this hyperboloid from the point x = y = z = t = u = 0 on the four-dimensional space u = R has the co-ordinates (ξ, η, ζ, τ). […]

This projection is limited by the 'hyperbola'

$$(b)R^2 + \xi^2 + \eta^2 + \zeta^2 - \tau^2 = 0$$

[…] which is the projection of the points at infinity on the hyperboloid (a). *The part of u = R which is outside the hyperbola (b) is the projection of the (two-bladed) hyperboloid which is conjugated to (a).* It will be seen from [De Sitter's line element] that on the limiting 'hyperbola' (b) all $g_{\mu\nu}$ become infinite" [my emphasis].

And De Sitter emphasized the specific difference between the two systems even after being transformed to new forms:[49]

"In both systems A and B it is always possible, at every point of the four-dimensional time-space, to find systems of reference in which the $g_{\mu\nu}$ depend only on one space-variable (the 'radius-vector'), and not on the 'time'. In the system A the 'time' of these systems of reference is the same always and everywhere, in B it is not".

In **June 20 1917** De Sitter compared between both models by using spherical polar coordinates, r, $\psi$, $\theta$.[50] He demonstrated that, in a *stationary state*, and if all matter is at rest without any stresses and pressures, then his system is legitimate exactly as Einstein's.

Suppose that $\rho_0$ is the average density of Einstein's "world-matter" and $\rho_1$ is "ordinary matter" (stars, sun, etc), and we define in this case: $\rho = \rho_0 + \rho_1$ ($\rho_1 < \rho_0$). In Einstein's world the stress energy tensor is $T_{\mu\nu} = 0$ with the exception of $T_{44} = g_{44}\rho$.
We neglect gravitation, and consider only the inertial field. We thus neglect ordinary matter $\rho_1$, and take $\rho_0$ constant. The field equations (2) then become: [51]

$$G_{ij} - g_{ij}\lambda - \frac{1}{2} g_{ij}\rho\kappa_0 = 0$$

$$G_{44} - g_{44}\lambda = -\frac{1}{2} g_{44}\kappa\rho_0$$

These can thus be satisfied by *two different systems* of $g_{\mu\nu}$:

**Solution A**: Einstein's static solution, the line element:

$$ds^2 = -dr^2 - R^2 \sin^2\frac{r}{R}[d\psi^2 + \sin^2\psi d\theta^2] + c^2 dt^2,$$

if (3) applies.

**Solution B**: De Sitter's static solution, the line element:

$$ds^2 = -dr^2 - R^2 \sin^2\frac{r}{R}[d\psi^2 + \sin^2\psi d\theta^2] + \cos^2\frac{r}{R}c^2 dt^2,$$

if

$$\rho_0 = 0, \lambda = \frac{3}{R^2}$$

applies.

The three-dimensional line element is in the two systems A and B:

$$(4)\ d\sigma^2 = dr^2 + R^2 \sin^2\frac{r}{R}[d\psi^2 + \sin^2\psi d\theta^2].$$

If $R^2$ is positive and finite, this is the line element of a three-dimensional space with a constant positive curvature.

And there is (**solution C**) the line element of special relativity with $\rho_0 = 0$, $\lambda = 0$:

$$ds^2 = -dr^2 - r^2[d\psi^2 + \sin^2\psi \, d\theta^2] + c^2 dt^2.$$

In solutions A and B three-dimensional space has constant positive curvature, and in C it is Euclidean. In addition, in solution A there is a world-matter, and in B and C we have $\rho_0 = 0$, the hypothetical world-matter does not exist.[52]

Now "The system B is the four-dimensional analogy of the three-dimensional space of the system A". We obtain, the set (1A) with the diagonal values (0, 0, 0, 1) and the set (1B) with the diagonal values (0, 0, 0, 0).

And this appears to be again the main point brought out by De Sitter: "In B and C the time is entirely relative, and completely equivalent to the other three coordinates". Since Einstein's whole world A is filled homogenously with matter, we have the $g_{44} = 1$ for all values of the four-coordinates, and at infinity t' = t, and thus system A introduces "a quasi-absolute time". And in system C we indeed have no relativity of inertia at all. "The set (1A) is invariant for all transformations for which (at infinity t' = t; the set (1B) is invariant for *all* transformations. It thus seems that the system A only satisfies the mathematical postulate of relativity if the latter is applied to three-dimensional space only […] The world-matter thus takes the place of the absolute space in Newton's theory, or of the 'inertial system'".[53] In the case of Einstein's solution this is a fundamental disturbing and painful feature.

De Sitter then thought it was certainly proper to assert: "It cannot be denied that the introduction of the constant $\lambda$, which distinguishes the systems A and B from C, is somewhat artificial, and detracts from the simplicity and elegance of the original theory of 1915, one of whose great charms was that it embraced so much without introducing any new empirical constant".[54]

**By June 1917**, De Sitter become aware of the experimental work by Vesto Melvin Slipher possessing the radial velocities of twenty-five spiral nebulae; yet in 1917 De Sitter knew of only three of them. On **April 13, 1917** Slipher contemplated the existence of possible red shifts from these nebulae: "In Table I. are given the velocities for the twenty-five spiral nebulae thus far observed. In the first column is the new general Catalogue number of the nebula and in the second the velocity. The plus sign denotes the nebula is receding, the minus sign that it is approaching". Slipher observed red shifts and interpreted these as velocity shifts: "Referring to the table of velocities again: the average velocity 570 km is about thirty times the average velocity of the stars. […] The mean of the velocities with regard to sign is positive, implying the nebulae are receding with a velocity of nearly 500 km".[55]

Considering the practicability of his universe, in a meeting of **June 17, 1917** De Sitter proposed a redshift effect in his hyperboloid world.[56]

"In the system *A* $g_{44}$ is constant, in *B* $g_{44}$ diminishes with increasing r". De Sitter referred to the temporal member in the line element of system B:

(5) $\cos^2 \dfrac{r}{R} c^2 dt^2$

"Consequently in *B* the lines is the spectra of very distant objects must appear displaced towards the red. This displacement by the inertial field is superposed on the displacement produced by the gravitational field of the stars themselves. It is well known that the Helium-stars show a systematic displacement corresponding to a radial velocity of + 4.3 Km/sec. If we assume that about 1/8 of this is due to the gravitational field of the stars themselves, then there remains for the displacement by the inertial field about 3 Km/sec. We should thus have, at the average distance of the Helium stars

$f$ = 1 − 2 · $10^{-5}$ = $\cos^2$(r/R).

If for this average distance we take r = 3 · $10^7$ […], this gives R = 2/3 · $10^{10}$. […] Lately some radial velocities of nebulae have been observed, which are very large; of the order of 1000 Km/sec. If we take 600 Km/sec., and explain this as a displacement towards the red produced by the inertial field, we should, with the above value of R, find for the distance of these nebulae r = 4 · $10^8$ = 2000 parsecs. It is probable that the real distance is much larger".

But this is what happened, Einstein was hard to give up Mach's ideas; instead he suggested that De Sitter's model was actually not-matter free, and Einstein could not possibly imagine another option to his own cosmological solution. It was this that led Einstein to quickly see the need to define the principles on which general relativity is based. In his **March 14, 1918** paper "Principles of the General Theory of Relativity", Einstein wrote that his theory rests on three principles, which are not independent of each other. He formulated the principle of relativity in terms of the Point Coincidence Argument. The three principles are: [57]

"a) *Relativity Principle*: The laws of nature are merely statements about space-time coincidences; they therefore find their only natural expressions in generally covariant equations.

"b) Equivalence Principle: Inertia and weight are identical in nature. It follows necessarily from this and from the result of the special theory of relativity that the symmetric 'fundamental tensor' [$g_{\mu\nu}$] determines the metrical properties of space, the inertial behavior of bodies in it, as well as gravitational effects. We shall denote the state of space described by the fundamental tensor as the 'G-field'."

"c) *Mach's Principle*[1)]: The G field is *completely* determined by the masses of the bodies. Since mass and energy are identical in accordance with the results of the special theory of relativity and the energy is described formally by means of the symmetric energy tensor ($T_{\mu\nu}$), this means that the G-field is conditioned and determined by the energy tensor of the matter.

1) Hitherto I have not distinguished between principles (a) and (c), and this was confusing. I have chosen the name 'Mach's principle' because this principle has the significance of a generalization of Mach's requirement that inertia should be derived from an interaction of bodies."

We can in fact say that Einstein was using Mach's Principle years before he coined the name "Mach's Principle" in 1918. Actually Einstein had worked hard to save his cosmological model over the previous year, so that for a while he truly believed in c) the Mach-Einstein principle.

In 1920, in a draft to a *Nature* paper Einstein explained that Mach later criticized Newton's mechanics; "but he was (after Newton) the first to vividly feel and clearly illuminate the epistemological weakness of classical mechanics". However, the natural equality between inertia and gravitation "remained hidden to Mach" [blieb Mach verborgen].[58] Actually we have noticed that it was Einstein who had demarcated between Mach's principle and the equivalence principle (inertia and weight are identical in nature) in his 1918 paper;[59] and therefore, the foregoing remarks applied to Einstein. The demarcation concerned "remained hidden" to Einstein until then and not to Mach. Turning again to problems in Newtonian mechanics, Einstein made an explicit use of the words "epistemological weakness" [erkenntnistheoretischen Schwächen] or "epistemological defect" of Newton's mechanics. Did Mach ever talk of "epistemological weakness"?

On **July 22, 1917** Einstein expressed doubts about the De Sitter matter-free universe, and concluded that De Sitter's assertions are incontestable; this led Einstein to submit another paper on **March 21, 1918** under the title: "Critical Comments on the Solution of the Gravitational Field Equations Given by Mr. De Sitter".[60]

It now seems to Einstein that De Sitter's model should have internal contradictions and something in his line element might be badly mistaken. What triggers Einstein to search for matter in De Sitter's universe is the temporal component of the static De Sitter metric (5):

$$g_{44} = \cos^2 \frac{r}{R} c^2$$

This component is variable: for r = 0, $g_{44}$ = 1 For r = ($\pi$/2)R, $g_{44}$ = 0, and it changes from 1 to 0. According to general relativity, the closer clocks are to a material source, the more slowly they run.

Einstein thus reasoned that, clocks slowed down as they approached r = ($\pi$/2)R, and all matter of De Sitter's world was concentrated in this "equator" at r = ($\pi$/2)R.

Einstein concluded that De Sitter's solution contained an intrinsic singularity indicating there is hidden matter in the surface $r = (\pi/2)R$: "The De Sitter system does in no way correspond to the case of matter-free world, but rather to a world, the material content of which is concentrated in the surface $r = (\pi/2)R$; and this could perhaps be proven by considering the limit of a spatial matter distribution turning into a surface distribution".[61]

De Sitter, and as we shall see also others, explained that this singularity was an artifact of the static form. The equator for an observer is a hypersurface in space-time which divides all events into two classes: those that have been, are, or will be observable by this observer, and those that are forever outside the observer's power of observation.[62] But it was unfortunate that Einstein was not persuaded and he repeatedly said that De Sitter's world still had matter, a matter-equator.

In **May 1920**, Arthur Stanley Eddington satisfied the curiosity of the general reader as well as the needs of the serious student by publishing his then "latest" book, *Space, Time, and Gravitation*, which was "excellently adapted to serve both classes".[63] Eddington was known as "the foremost champion of *Einsteinismus* in English". Attempting to explain the standpoint which Einstein took in his polemic with De Sitter, Eddington borrowed from traditional British literature nursery stories such as *Alice in Wonderland*: [64]

"Spherical space-time, that is to say a four-dimensional continuum of space and imaginary time forming the surface of a sphere in five dimensions, has been investigated by Prof, de Sitter. If real time is used the world is spherical in its space dimensions, but open towards plus and minus infinity in its time dimension, like an hyperboloid. This happily relieves us of the necessity of supposing that as we progress in time we shall ultimately come back to the instant we started from! [spherical world]

History never repeats itself. But in the space dimensions [line element (4)] we should, if we went on, ultimately come back to the starting point. This would have interesting physical results, and we shall see presently that Einstein has a theory of the world in which the return can actually happen; but in de Sitter's theory it is rather an abstraction, because, as he says, 'all the paradoxical phenomena can only happen after the end or before the beginning of eternity.'

The reason is this. Owing to curvature in the time dimension [the variable temporal member (5) in the static line element], as we examine the condition of things further and further from our starting point, our time begins to run faster and faster, or to put it another way natural phenomena and natural clocks slow down. The condition becomes like that described in Mr H. G. Wells's story 'The new accelerator.'

When we reach half-way to the antipodal point, time stands still. Like the Mad Hatter's tea party, it is always 6 o'clock; and nothing whatever can happen however long we wait".

This was surely true in the static form, in the time dimension things depended on position; and Einstein claimed that clocks slowed down as we approached the surface r = (π/2)R. This slowing down of clocks apparently indicated that all the matter of De Sitter's world must be concentrated in this equator. *Einstein thought that no choice of coordinates could remove this singularity*; and he was certainly wrong.

Eddington's biographer, Allie (Alice) Vibert Douglas from Queens University (Kingston, Ontario) visited Einstein in 1954, a year before his death. She spoke with Einstein forty minutes about cosmology and relativity: [65]

"In early January 1954 I visited Dr. Albert Einstein as a direct consequence of a research into the history of cosmological thought during the years immediately following the publication in 1915 of his General Theory of Relativity. I wanted to weigh the contributions of the three immediate contributors to the development of this theory, namely de Sitter, Weyl and Eddington, and especially to see Eddington's contribution through Einstein's eyes. […]

He came directly to the point of my visit and paid a striking tribute to the English astronomer, Sir Arthur Eddington, who was the first and greatest interpreter of the theory of general relativity to the English-speaking world. He spoke of the literary value, the beauty and brilliance of Eddington's writing in those books aimed at giving to the intelligent lay reader at least some understanding, some insight into the significance of the new scientific ideas – but with a smile he added that a scientist is mistaken if he thinks he is making the layman understand; a scientist should not attempt to popularize his theories, if he does 'he is a fakir – it is the duty of a scientist to remain obscure'. I said I could not agree, that the scientist had a duty to try to educate the public at least to an appreciation of what the scientist is attempting to do; but Dr. Einstein shook his head".

According to Vibert Douglas, Einstein did not quite like Eddington's popular writings. Alternatively, the reason could perhaps be that Einstein did not like Eddington's cosmological model (to be described later), but Einstein very likely disliked Eddington's criticism of his standpoint in the successful book, *Space, Time, and Gravitation*.[66]

Back in **May-June 1918**, Felix Klein demonstrated to Einstein that the equator in the static form of the De Sitter solution is an artifact of the way in which the time coordinate is introduced. Einstein failed to appreciate that Klein's analysis of the De Sitter solution showed that the singularity at the equator can be transformed away and does not indicate the presence of matter after all. In his response, Einstein simply reiterated the argument of his critical note on the De Sitter solution, for which Weyl in **May 1918**, he thought, had just provided new support.[67] Weyl concluded that the possibility of an empty world contradicts the laws of nature which we adopt as valid, because at least at the horizon there must have existed masses. Weyl's position corresponded exactly to Einstein's when he criticized De Sitter's solution. This reveals the "influence of an authority in physics even on first-rate mathematicians".[68]

Klein then brought as precise an explanation as possible for the coordinate singularity in De Sitter's model; he tried to persuade Einstein that it was an oversight which occurred in the static form, and it seemed that some of the difficulties encountered in De Sitter's model could be overcome if we recognized the possibility that the singularity was indeed a result of the transition to the static form. Klein explained why it seemed as if the surface r = (π/2)R could never be reached: the coordinate system in which the De Sitter world was static covered only part of the entire De Sitter space-time, and the surface r = (π/2)R was not included. Klein had shown that in the static form of the De Sitter solution, the time coordinate breaks down on the equator.

Eventually Einstein understood that there was no doubt about the mathematical existence of De Sitter's solution. He thus wrote back to Klein and admitted that the De Sitter solution was matter-free and fully regular. In the circumstances, the De Sitter solution was indeed a counterexample to Mach's principle. With the help of Klein Einstein understood (though not immediately) that, because of the mapping of this model into a static form, the De Sitter solution seemed to him not to correspond to a world that was both free of singularities and free of world-matter.

In those days Einstein's colleagues and friends could have thought, had he been less stuck to Mach's principle, he would have been less embarrassed. It was indeed now almost inevitable for Einstein to agree that De Sitter's solution was a matter-free solution, but he still did not accept De Sitter's world as a possible physical cosmological model. From his standpoint it was not just a matter of mathematical convenience as to whether we choose to map De Sitter's model to a static form. Einstein held that the representation of the world itself would be different, i.e. any acceptable cosmological model would have to be static.[69]

Finally, William McCrea wrote that De Sitter's model was less serious threat to Einstein's position than at first appeared to be the case, because indeed De Sitter's model was not properly to be regarded as static. It could be given an apparently static form only as a result of *a mathematical accident*. So long as astronomers kept to the idea that the universe as a whole has to be static, Einstein's model was the only known theoretical model satisfying general relativity field equations.[70]

In retrospect, this discussion on singularities proved to be of much significance to the consideration of the existence of the big bang: "the theory from which it all began in 1917 – particularly in regard to the occurrence and nature of singularities". The big bang in a cosmological model is associated with a singularity in its space-time. The occurrence of some sort of singularity has been shown to be inevitable in any space-time that is likely to be of physical interest.[71]

In the first edition of *Raum-Zeit-Materie*, chapter §33, Weyl attempted to establish a field theory of gravitation within the frame of Einstein's general relativity. He wished to deduce the electromagnetic equations from a generalization of Riemannian geometry. Within this framework Weyl explained that Einstein found himself constrained to assume that the world is closed with respect to space; for in this case the boundary conditions are naturally dropped. Weyl showed that "*In any case, we see that the differential equations of the field contain the physical laws of nature in their complete form*, and that there cannot be a further limitation due to boundary conditions at spatial infinity". Weyl adopted Einstein's cosmological model. But he claimed that in terms of the above explanation of deduction of electromagnetism from gravitation, and assuming his above suggestion, we could let the cosmological constant be as small as possible.[72]

Einstein was inspired by Weyl's above idea. At this stage Einstein resented the fact that his cosmological term was a "kosmologischen Zusatzglied". In **April 1919** Einstein published a paper "Do gravitational fields play an essential part in the structure of the elementary particles of matter?" It was tempting for him to have his cake and eat it too. Einstein was uneasy about introducing the cosmological term into his field equations from the beginning. He now reflected upon a topic slightly different, inspired by Weyl and others (Gustav Mie and David Hilbert), gravitation and electromagnetic fields; but when he worked on this problem he realized like Weyl that this was somehow related to the closed universe and the cosmological problem; he found such a marvelous idea which was significant for getting rid of the horrifying ad-hoc character of the cosmological "glied".[73]

Einstein exchanged his field equations:

$$(1)\ R_{ik} - \frac{1}{2} g_{ik} \mathrm{R} = -\kappa \mathrm{T}_{ik}$$

($R_{ik}$ the Ricci curvature tensor, R the scalar curvature, $g_{ik}$ the metric tensor, and $T_{ik}$ the stress-energy tensor) for equations:

$$(1a)\ R_{ik} - \frac{1}{4} g_{ik} \mathrm{R} = -\kappa \mathrm{T}_{ik}$$

with the electromagnetic field tensor $T_{ik}$ as a source.

This new formulation is now called "trace-free" Einstein field equations.

Einstein showed that we could start from the above unmodified field equations (1) plus an additional cosmological term $\lambda$:

$$R_{ik} - \lambda g_{ik} = -\kappa \left( \mathrm{T}_{ik} - \frac{1}{2} g_{ik} T \right)$$

subtract the scalar equation multiplied by 1/2 and obtain:

$$[(9a)] \left( R_{ik} - \frac{1}{2} g_{ik} \mathrm{R} \right) + \lambda g_{ik} = -\kappa T_{ik}$$

It seemed to Einstein remarkable that "the new formulation has this great advantage that, the quantity λ appears in the fundamental equations as a constant of integration and no longer as a universal constant peculiar to the fundamental law".[74]

Einstein found that we could start from equations (1a) that had no additional cosmological term λ, and using λ as a constant of integration, we obtained (9a), the field equations with an additional cosmological term λ.

Einstein began spending much effort at getting rid of the cosmological constant; he initiated no fundamentally new model for the universe; at this stage he was not thinking about the reality of non-static solutions, but he intended a program to regain a natural and simple formulation for the field equations.

Nevertheless, in 1920 Einstein thought in terms of "Mach's Ether". The "world-matter" of Einstein's world was equivalent to an ether, to "Mach's ether", a Machian substance that was needed as a carrier of the effects of inertia. In an address he gave on **May 5, 1920** in the University of Leiden, Einstein thought his "world-matter", or new ether, was absolutely essential: "since the modern physicist does not believe that he may accept action at a distance, he comes back once more, if he follows Mach, to the ether, which has to serve as medium for the effects of inertia".[75]

Einstein explained this further: [76]

"As to the role which the new ether is to play in the physical world picture of the future, we are not yet clear. We know that it determines the metric relations in the space-time continuum, e.g. the configuration possibilities of solid bodies as well as the gravitational fields; but we do not know whether it has an essential share in the structure of the electrical elementary particles constituting matter. Nor do we know whether it is only in the vicinity of ponderable masses that its structure differs significantly from that of the Lorentzian ether; whether the geometry of spaces of cosmic extent is approximately Euclidean. But we can argue on the basis of the relativistic equations of gravitation that there must be a deviation from the Euclidean behavior, with spaces of cosmic order of magnitude, if there exists a positive mean density of matter, no matter how small, in the universe. In this case the world must necessarily be spatially closed and of finite size, and its magnitude being determined by the value of that mean density".

It would have been very natural in 1920 for the Mach-Einstein's principle to be very helpful for the purpose of inventing "Mach's ether". Einstein still had strong predilection for Mach's ideas, but all the same he brought back nothing less than the ether.

Popularly thinking in terms of general-relativity's space-time, fairly obviously there is so-called space-time "Mach's ether". But the notion "Mach's ether" instead of De Sitter's "world-matter" did not so much hold a significant role in Einstein's thought and did not last long. More importantly, in the year 1921 Einstein again lectured on what was generally seen as his notorious cosmological model and even summarized his ideas from his 1917 cosmology paper.

It happens, however, that the "Mach's ether" address as stated was given in the University of Leiden. If anyone in 1921 was concerned about an ether, it seems obvious that this would be Lorentz, Einstein's life-long friend and colleague from Leiden. Perhaps Einstein invoked Mach's ether" simply because Lorentz was stuck to the ether and Einstein always was eager to respect him. To some extent, Lorentz was unwilling to accept relativity as is. Further, since it was difficult to explain to a general audience the new features of general relativity and the cosmological model apart from their general setting, Einstein's new Machian ether appeared to be justified.

### Einstein visits America and lectures on his 1917 cosmological model

In **April-May 1921** Einstein joined Chaim Weizmann's tour to the United States to gain support among American Jewry for the Zionist cause. His role was to raise funds for the establishment of Hebrew University of Jerusalem.

On **April 29, 1921** Einstein visited City College in New York. Professor Morris R. Cohen wrote a review of Einstein's lectures in which he mentioned Einstein's cosmological model: [77]

"In the concluding portion of his lecture, Professor Einstein dealt briefly with the considerations which lead him to reject the idea of a universe containing a finite amount of matter in an infinite space. An infinite amount of matter seems to be incompatible with the known behavior of bodies. We are, therefore, forced to conclude that space is finite".

Einstein then went to Princeton. Princeton University has arranged five lectures on the theory of relativity on the afternoon from **May 9 to 13, 1921**, the subject of these lectures, which were delivered *in German*, were special relativity, general relativity, "Generalities on the Theory of Relativity", and "Cosmological Speculations".

On **April 30, 1921** invitations were sent out to more than 600 college and university presidents to attend this series of lectures. The last lecture given the afternoon of **May 13, 1921** dealt with general relativity and cosmology. Einstein talked in McCosh 50 hall at 4:15 in the afternoon. The lecture took the technical side of general relativity, comparing it to the lecture dealing with special relativity discussed the day before. Reporters wrote that, "The blackboard was covered with abstract diagrams which only a trained mathematician could follow". The "Cosmological Speculations" ended Einstein series of lectures and terminated his visit at Princeton. [78]

According to a *New York Times* Reporter, in the final May 13, 1921 lecture "Cosmological Speculations", Einstein explained his concepts of "finite and yet unlimited universe". And the reporter added: "Just what the size of the universe is he said could not be determined at present, because it is first necessary to know the mean density of matter in it, and this at present is a quantity of which there is no knowledge". The reporter explained to the readers of the *New York Times*: "Professor Einstein's idea of the finite universe is that of a spherical universe of finite extent, but infinite because of its curved nature […] He conceives the universe as being bent back upon itself much as the mythical snake which swallows its tail, although, of course, there is no way of making a graph of what is a mathematical abstraction".[79]

The lectures were later published in a book form by the Princeton University Press. On **May 11, 1921** Einstein, "has agreed not to authorize the publication of his lectures in the United States by anyone else, so that this volume will contain not only the latest exposition of the Theory of relativity, but it will be the only one authorized by the famous scientist during his visit to this country. A German stenographer is taking notes of the lectures as they are delivered. The plan of procedure is to have her write her notes out in German and then Professor Edwin P [Plimpton]. Adams of the Department of Physics will go over them and check up those scientific portions which may have caused trouble. After Professor Adams has completed this part of the work, the lectures will be submitted to Professor Einstein for revision and final approval. When he had returned them, they will be translated into English and published". After each address delivered by Einstein in German Edwin P. Adams made a résumé.[80]

Adams is reported to have said in a summary of Einstein's lecture "Cosmological Speculations":[81]

"It is a remarkable fact that the general theory of relativity, built up as it is from physical considerations resulting from experiments on the earth, should have anything to say concerning the problem of the universe as a whole. It has generally been thought that the universe is infinite in extent. Telescopes of increasing power have brought more and more distant stars to our vision".[82]

And then Adams repeats Einstein's explanation from his 1917 Cosmological paper.[83] As already stated, Einstein had started the discussion of the cosmological model with the Newtonian world. Newtonian classical universe was plugged with the problem which was mentioned by Seeliger and Neumann for Newtonian gravitational theory, a static infinite universe filled with matter needed a repulsive force that varied with the distance.[84] According to the Adams summary of Einstein's lecture of **May 13, 1921**:[85]

"If we imagine a sphere of radius very large compared to the mean distance between the stars, our first view is that as we increase the radius of the sphere more and more a definite density of matter in the universe is approached. The astronomer Seeliger first showed that such a view is definitely opposed to the Newtonian law of gravitation, for this view immediately leads to the result that the gravitational field would also increase beyond all limits as we go out toward infinity, and this would mean that the stellar velocities would necessarily increase beyond all limits".

Adams then summarizes: [86]

"Thus on the basis of Newton's theory we should have to conclude that the mean density of matter in the universe is zero. This could only be attained by assuming that the universe is an island floating in infinite space free from matter. But this view is wholly unsatisfactory, and Seeliger attempted to reconcile an infinite universe with finite density by assuming that matter of negative density is present in the universe. This assumption involves departure from Newton's law of gravitation, but no other argument leads to a similar conclusion, and so this is not a satisfactory solution".

Adams then explains that Einstein made a slight modification to his general theory of relativity which does not change any of the other conclusions drawn from it.

In this published **1922** book, *The Meaning of Relativity*, in lecture III, Einstein (in Adams translation) explained again relativity of inertia, Mach's principle: [87]

"[…] it is contrary to the mode of thinking in science to conceive of a thing (the space-time continuum) which acts itself, but which cannot be acted upon. This is the reason why E. Mach was led to make the attempt to eliminate space as an active cause in the system of mechanics. According to him, a material particle does not move in unaccelerated motion relatively to space, but relatively to the centre of all the other masses in the universe; in this way the series of causes of mechanical phenomena was closed, in contrast to the mechanics of Newton and Galileo. In order to develop this idea within the limits of the modern theory of action through a medium, the properties of the space-time continuum which determine inertia must be regarded as field properties of space, analogous to the electromagnetic field. The concepts of classical mechanics afford no way of expressing this. For this reason Mach's attempt at a solution failed for the time being".

In lecture IV Einstein (in Adams translation) further explained why according to Mach's ideas one is obliged to choose his cosmological model and abandon the boundary conditions: [88]

"If the universe were quasi-Euclidean, then Mach was wholly wrong in his thought that inertia, as well as gravitation, depends upon a kind of mutual action between bodies. For in this case, with a suitably selected system of co-ordinates, the $g_{\mu\nu}$ would be constant at infinity, as they are in the special theory of relativity, while within finite regions the $g_{\mu\nu}$ would differ from these constant values by small amounts only, with a suitable choice of co-ordinates, as a result of the influence of the masses in finite regions. The physical properties of space would not then be wholly independent, that is, uninfluenced by matter, but in the main they would be, and only in small measure, conditioned by matter. Such a dualistic conception is even in itself not satisfactory; there are, however, some important physical arguments against it, which we shall consider.

The hypothesis that the universe is infinite and Euclidean at infinity, is, from the relativistic point of view, a complicated hypothesis. In the language of the general theory of relativity it demands that the Riemann tensor of the fourth rank $R_{iklm}$ shall vanish at infinity, which furnishes twenty independent conditions, while only ten curvature components $R_{\mu\nu}$, enter into the laws of the gravitational field. It is certainly unsatisfactory to postulate such a far-reaching limitation without any physical basis for it".

So far as Mach's principle Einstein was not willing to give up: [89]

"But in the second place, the theory of relativity makes it appear probable that Mach was on the right road in his thought that inertia depends upon a mutual action of matter. For we shall show in the following that, according to our equations, inert masses do act upon each other in the sense of the relativity of inertia, even if only very feebly". And: "We must see in them a strong support for Mach's ideas as to the relativity of all inertial actions. If we think these ideas consistently through to the end we must expect the whole inertia, that is, the whole $g_{\mu\nu}$-field, to be determined by the matter of the universe, and not mainly by the boundary conditions at infinity".

He thought he had a very strong epistemological argument, Mach's principle, in regarding his theory: [90]

"Thus we may present the following arguments against the conception of a space-infinite, and for the conception of a space-bounded, universe:

1. From the standpoint of the theory of relativity, the condition for a closed surface is very much simpler than the corresponding boundary condition at infinity of the quasi-Euclidean structure of the universe."

The quasi-Euclidean structure of the universe is Minkowski space-time of the special theory relativity. In this case, the $g_{\mu\nu}$ would be constant at infinity, an empty world, exactly as they are in the special theory of relativity. Einstein could not accept this. He explained:

"2. The idea that Mach expressed, that inertia depends upon the mutual action of bodies, is contained, to a first approximation, in the equations of the theory of relativity; it follows from these equations that inertia depends, at least in part, upon mutual actions between masses. As it is an unsatisfactory assumption to make that inertia depends in part upon mutual actions, and in part upon an independent property of space, Mach's idea gains in probability. But this idea of Mach's corresponds only to a finite universe, bounded in space, and not to a quasi-Euclidean, infinite universe. From the standpoint of epistemology it is more satisfying to have the mechanical properties of space completely determined by matter, and this is the case only in a space-bounded universe.

3. An infinite universe is possible only if the mean density of matter in the universe vanishes. Although such an assumption is logically possible, it is less probable than the assumption that there is a finite mean density of matter in the universe". Einstein did not accept other models, including that of De Sitter, as a physical possibility.

Back to Einstein's stay in America; later Einstein travelled to Boston. Einstein began his second day of his Boston visit with a motor ride to Cambridge, were president of the Harvard University gave him a semi-formal reception. Einstein was much interested in the university, partly because it included the Harvard Astronomical Observatory. But he was to hurry back to Boston, because his schedule included two afternoon receptions, one of which was with Zionists,[91] and so he was again stuck with his good old cosmological model.

In passing, it may be remarked that **in 1922** Jacobus Cornelius Kapteyn (then at Mount Wilson Observatory), who promoted the Dutch school of astronomy, published a first attempt at a theory of the distribution of masses, forces, and velocities in the stellar system. He noted: "It is incidentally suggested that when the theory is perfected it may be possible to determine *the amount of dark matter* from its gravitational effect".[92]

And Kapteyn further explained: [93]

"*Dark Matter*. It is important to note that what has here been determined is the total mass within a definite volume, divided by the number of luminous stars. I will call this mass the average effective mass of the stars. It has been possible to include the luminous stars completely owing to the assumption that at present we know the luminosity-curve over so large a part of its course that further extrapolation seems allowable.

Now suppose that in a volume of space containing $l$ luminous stars there be dark matter with an aggregate mass equal to $Kl$ average luminous stars; then, evidently the effective mass equals *(l + K)* X average mass of a luminous star.

We therefore have the means of estimating the mass of dark matter in the universe".

What a triumph this could be to the Mach-Einstein principle and to Einstein's cosmological constant in the wonderland of science at those days.

### Redshift and non-static effects in De Sitter world

For some years subsequently, observational astronomers did not accept with much faith the cosmological models of Einstein and De Sitter. For an astronomer to do so, an acceptance of General relativity was a necessary condition, and many of them were skeptical towards the theory. Even after the results of the 1919 British eclipse expeditions had apparently vindicated predictions of the theory, a majority of astronomers felt uneasy with the mathematical tools of the theory. [94]

Nevertheless a quite different path had been taken by Cosmologists' extensive studies of De Sitter's solution to the general relativity field equations. They took the stance that perhaps De Sitter's world was preferable, and they proceeded to study De Sitter's model. Eddington and Weyl inferred that Slipher's observations could be explained using De Sitter's model; the ideas underlying their suggestion were the redshift effect that had already been discovered by De Sitter in 1917. Weyl derived a relation representing observable quantities – mainly a theoretical relation between redshift and distance in the used model of De Sitter's world; and applied this to Slipher's recent observational data. In applying redshift relations (and recession in De Sitter model) there was a problem, because Slipher sometimes obtained different numerical values in his observations and extra details unknown at the time to cosmologists; and thus they could not bring his results into full agreement with De Sitter's model.

In **December 1920** Hermann Weyl finished to write the fourth edition of *Raum-Zeit-Materie* (*Space-Time-Matter*). He adhered to the present model of Einstein and did not accept De Sitter's model. He claimed that De Sitter's model contained a singularity and thus a hypersurface, a mass-horizon.

Weyl repeated the central idea presented in the previous editions:[95]

"*In any case, we see that the differential equations of the field contain the physical laws of nature in their complete form*, and that there cannot be a further limitation due to boundary conditions at spatial infinity.

Einstein, arguing from cosmological considerations of the interconnection of the world as a whole came to the conclusion that the world is finite in space. Just as in the Newtonian theory of gravitation the law of contiguous action expressed in Poisson's equation entails the Newtonian law of attraction only if the condition that the gravitational potential vanishes at infinity is added, so Einstein in his theory seeks to supplement the differential equations by introducing boundary conditions at spatial infinity. To overcome the difficulty of formulating conditions of a general invariant character, which are in agreement with astronomical facts, he finds himself constrained to assume that the world is closed with respect to space; for in this case the boundary conditions are naturally dropped".

But Weyl cannot admit the cogency of this deduction, since "the differential equations in themselves, without boundary conditions, contain the physical laws of nature in an unabbreviated form excluding every ambiguity".

Although Weyl used different differential equations than Einstein's, he imposed apparently the same general idea. Weyl presented Olbers' paradox and then concluded: ("*Der Raum stellt sich als geschlossen und daher endlich heraus*") "*space* *is found to be closed and hence finite.* If this were not the case, it would scarcely be possible to imagine how a state of static equilibrium could come about".

So long as Weyl kept to the cosmological constant and to Einstein's basic tenets, his conclusions were that found in Einstein's work of 1917.

Einstein indeed exerted the greatest influence upon Weyl's ideas, so much that even by **December 1920** Weyl still thought:[96]

"But since on the 'greatest sphere' $x_4 = 0$, [of De Sitter's world] which may be designated as the equator or the space-horizon for that centre, $f = 0$, and hence the fundamental metric of the world becomes singular, we see that the possibility of a static empty world is contrary to the physical laws – that are here regarded as valid – at least at the horizon where there must be masses".

The change here Weyl envisaged in favor of electromagnetism and gravitation unified in some manner was not received very well by Eddington. On **February 19, 1921** he wrote: "H. Weyl has shown that, on removing a rather artificial restriction in Riemann's geometry, the expression for the metric includes also terms which are identical with the four potentials of the electromagnetic field. I believe that Weyl's geometry, far-reaching though it is, yet suffers from an unnecessary and harmful restriction; and it is the object of this paper to develop a still more general theory". Eddington was quite against "Weyl's generalized geometry", and he spoke of a "miracle" in Weyl's theory, and claimed to have provided calculations which provided "*precisely the same miracle* viewed from another standpoint".[97]

In **February 1921** *Nature* published a short paper by Einstein. The paper was a three pages abridged adaptation of the 1920 draft (and also an English translation of the original German by Robert W. Lawson). [98]

Einstein did not agree with Eddington's opinion (in *Space, Time, and Gravitation*) and at the end of his 1921 paper we find the following explanation:[99]

"A final question has reference to the cosmological problem. Is inertia to be traced to mutual action with distant masses? And connected with the latter: Is the spatial extent of the universe finite? It is here that my opinion differs from that of Eddington. With Mach, I feel that an affirmative answer is imperative, but for the time being nothing can be proved. Not until a dynamical investigation of the large systems of fixed stars has been performed from the point of view of the limits of validity of the Newtonian law of gravitation for the immense regions of space will it perhaps be possible to obtain eventually an exact basis for the solution of this fascinating question".

In **August 1922**, Eddington completed his book *The Mathematical Theory of Relativity*. At that time both Einstein's and De Sitter's cosmological models remained without empirical support. Thus a decision in favor of one or the other model was impossible on empirical grounds. Scholars though uncovered defects in each model. In raising the issue of the distinction between the two solutions, Eddington had taken a leading part in revealing the limitations of the original Einstein model. Eddington explained that as they stood, the "Two forms of the world have been suggested –

(1) Einstein's cylindrical world. Here the space-dimensions correspond to a sphere, but the time-dimension is uncurved.

(2) De Sitter's spherical world. Here all dimensions are spherical; but since it is imaginary time which is homogeneous with the space-coordinates, sections containing real time become hyperbolas instead of circles".

We cannot choose a universe mixing the two, or what is practically equivalent "It seems natural to regard de Sitter's and Einstein's forms as two limiting cases, the circumstances of the actual world being intermediate between them. De Sitter's empty world is obviously intended only as a limiting case; and the presence of stars and nebulae must modify it, if only slightly, in the direction of Einstein's solution.

Einstein's world containing masses far exceeding anything imagined by astronomers, might be regarded as the other extreme – a world containing as much matter as it can hold. This view denies any fundamental cleavage of the theory in regard to the two forms, regarding it as a mere accident, depending on the amount of matter which happens to have been created, whether de Sitter's or Einstein's form is the nearer approximation to the truth. But this compromise has been strongly challenged, as we shall see".[100]

Later, Eddington began his 1933 book, *The expanding Universe*, with the "The De Sitter spectral shift effect". In a truly ironical manner he implicitly admitted that there was something strange in the De Sitter model that was embraced by him:[101]

The first hint of an 'expanding universe' is contained in a paper published in November 1917 by Prof. W. de Sitter. Einstein's general theory of relativity had been published two years before, but it had not yet attained notoriety; it was not until the eclipse expeditions of 1919 obtained confirmation of its prediction of the bending of light that public interest was aroused. Meanwhile many investigators had been examining the various consequences of the new theory. Prominent among them was de Sitter who was interested especially in the astronomical consequences. In the course of a highly technical discussion he found that the relativity theory led to an expectation that *the most remote celestial objects would be moving away from us*, or at least that they would deceive the observer into thinking that they were moving away.

De Sitter was perhaps a tipster rather than a prophet; […] he suggested that we ought to keep a look out for the recession as a rather likely phenomenon".

Eddington spoke in favor of De Sitter's model:[102]

"We have now realized that the changelessness of de Sitter's universe was a mathematical fiction. Taken literally his formulae described a *completely empty* universe; but that was meant to be interpreted generously as signifying that the average density of matter in it, though not zero, was low enough to be neglected in calculating the forces controlling the system. It turned out, however, that the changelessness depended on there being literally no matter present. In fact the 'changeless universe' had been invented by the simple expedient of omitting to put into it anything that could exhibit change. We therefore no longer rank de Sitter's as a static universe; and Einstein's is the only form of material universe which is genuinely static or motionless".

Einstein's world was not only the static world, but also the only Machian world. Since the time Einstein and De Sitter debated on which model represented the actual universe, De Sitter described them both in a tricky manner in a 1930 meeting of the Royal Astronomical Society: "Einstein's solution gives a world full of matter, but no motion; mine gives a world full of motion, but no matter".[103] Eddington, attending that meeting, later adopted this phraseology in his 1933 book, *The expanding Universe*:[104]

"The situation has been summed up in the statement that Einstein's universe contains matter but no motion and de Sitter's contains motion but no matter. It is clear that the actual universe containing both matter and motion does not correspond exactly to either of these abstract models. The only question is, Which is the better choice for a first approximation? Shall we put a little motion into Einstein's world of inert matter, or shall we put a little matter into de Sitter's Primum Mobile?"

But if we are to make further progress in our understanding of the actual universe it appears, for the reasons stated, essential to study De Sitter's universe; Einstein's world is static and De Sitter's has been found to be globally non-static. Eddington explained his choice in De Sitter's world:[105]

"The clock-beats become longer and longer as we recede from the origin; in particular the vibrations of an atom become slower. Moreover we can detect by practical measurement this slowing down of atomic vibrations, because it is preserved in the transmission of the light to us. The coordinates [of De Sitter] form a statical system, the velocity of light being independent of t; hence the light-pulses are all delayed in transmission by the same "time" and reach us at the same intervals of t as they were emitted. Spectral lines emanating from distant sources at rest should consequently appear displaced towards the red. At the 'horizon' $1/2 \pi R$ [equator] any finite value of ds corresponds to an infinite dt. It takes an infinite 'time' for anything to happen. All the processes of nature have come to a standstill so far as the observer at the origin can have evidence of them.

But we must recall that by the symmetry of the original formula, any point of space and time could be chosen as origin with similar results. Thus there can be no actual difference in the natural phenomena at the horizon and at the origin. The observer on the horizon does not perceive the stoppage – in fact he has a horizon of his own at a distance 1/2 πR where things appear to him to have come to a standstill".

De Sitter's model was apparently preferable because of recessional velocities of Nebulae measured by Slipher. It turned out that De Sitter's model revealed non-static properties that Eddington inclined to accept it: [106]

"Thus [in De Sitter's world] a particle at rest will not remain at rest unless it is at the origin; but will be repelled from the origin with an acceleration increasing with the distance. A number of particles initially at rest will tend to scatter, unless their mutual gravitation is sufficient to overcome this tendency.

It can easily be verified that there is no such tendency in Einstein's world. A particle placed anywhere will remain at rest. This indeed is necessary for the self-consistency of Einstein's solution, for he requires the world to be filled with matter having negligible velocity. It is sometimes urged against de Sitter's world that it becomes non-statical as soon as any matter is inserted in it. But this property is perhaps rather in favour of de Sitter's theory than against it".

And this apparent alteration or change in De Sitter's universe was a strange property that led cosmologists in the 1920s to adhere to this model, because:

"One of the most perplexing problems of cosmogony is the great speed of the spiral nebulae. Their radial velocities average about 600 km. per sec. and there is a great preponderance of velocities of recession from the solar system.

It is usually supposed that these are the most remote objects known (though this view is opposed by some authorities), so that here if anywhere we might look for effects due to a general curvature of the world. De Sitter's theory gives a double explanation of this motion of recession; first, there is the general tendency to scatter […]; second, there is the general displacement of spectral lines to the red [redshift] in distant objects due to the slowing down of atomic vibrations which would be erroneously interpreted as a motion of recession".

Eddington now proceeded to observations: "The most extensive measurements of radial velocities of spiral nebulae have been made by Prof. V. M. Slipher at the Lowell Observatory", according to which the nebulae exhibit velocities of recession from an observer.  "He has kindly prepared for me the following table, containing many unpublished results"; but De Sitter's model was not able to account for all of Slipher's findings and the difficulties were not overcome: "Even if these also show preponderance of receding velocities the cosmological difficulty is perhaps not entirely removed by the De Sitter's theory" sufficiently accurately.[107]

As regards application to the actual universe, in 1922 Weyl also turned in favor of De Sitter's universe and rejected the world of Einstein. In the 1923 Appendix III of the fifth edition of the book *Raum, Zeit, Materie*, Weyl combined De Sitter's world, the four-dimensional "sphere" hyperboloid, and the hypothesis, which became known as "Weyl's Principle", that stars lie on a "bundle" of geodesics that diverge from a common event in the past. Such an assumption was necessary to derive the cosmological redshift (dependence upon distance of the redshift of radiation) in de Sitter's model. [108]

Before **autumn 1922**,[109] Weyl found that spectral lines show redshift to a first approximation proportional to their distances in De Sitter's world. Weyl's considerations were suggested in connection with Slipher's observed apparent recession of the nebulae. A particular consequence of the work of Weyl in the fifth edition of *Space, Time, Matter*, was its indication that Einstein's cosmological term is related to a non-static element in De Sitter's world. Weyl adopted the cosmological constant which was seen as having the role of a cosmic repulsion term, forcing the world-lines of stars to recede with time, and it was connected to the redshifts of nebulae. The cosmological constant bears then a simple interpretation in terms of Weyl's principle: the world-lines of the stars in De Sitter's universe diverge from a single "bundle" of infinite geodesics. They diverge from an event in the past to a universal future, and this tendency "reflects a demonstration of the cosmological term".[110]

Weyl begins with two aspects of his new suggestion. In a complete cosmology supplementary assumptions must be added, which determine whether the entire De Sitter hyperboloid, or which part of it, corresponds to the real world, and another assumption about the motion of the stars by which infinite geodesic world-lines are set off from the manifold of all these lines (Weyl's principle): [111]

"The geodesics are cut out of the sphere [hyperboloid] by the two-dimensional planes passing through the origin in the five-dimensional space […]. The null cones opening into the future, which issue from all the points of such a geodesic with time-like direction, from the world-line of a star, fill a region of the world which I shall call *the range of influence of the star*. It is highly remarkable feature of the De Sitter cosmology that this range of influence covers only half the hyperboloid (while it coincides with the entire 'plane' in the special theory of relativity). […] (The sector represented by the corresponding static coordinates is again only part of the range of influence, and more precisely that part which is accessible to observation from [a star A]. There are [infinite] stars or geodesics to which the same range of influence belongs as to the arbitrary chosen star A; they form a *system that has been causally interconnected since eternity*. Stars that do not belong to it lie beyond the range of influence of A during their history. On the other hand, it is true that if A' is a star of the system, A ceases to act upon A' from a certain moment of its history on, even thought conversely A' remains in the range of influence of A during its entire history. Therefore the stars of the system may be described as stars 'of common origin' but the common origin lies in an infinitely distant past. *Our assumption is that in the undisturbed state the stars form such a system of common origin*".

Thus the findings bear a simple interpretation in terms of a "common origin" in De Sitter's universe, so that the world-lines cover only half of the hyperboloid which represents their common sphere of influence as the real world. "Only when referring to the entire hyperboloid is it appropriate to describe an infinite number of world-lines of the system of stars as one that concentrates on an infinitely small part of the total extent of the hyperboloid towards the infinitely distant past, while it spreads over it more and more towards the infinitely distant future". But only half which is covered by the world-lines has real significance.[112]

Weyl calculated the dependence upon distance of the redshift of the spectral lines for De Sitter space. He arrived at the following expression for the displacement of the spectral lines of the nebulae: [113]

$d v / v = 1 + \tan (r / R)$.

with r the measured distance of the star in the static space at the moment that the observation takes place, and R is the constant curvature of De Sitter's world.

This reduces for small r/R to: [114]

$d v / v \approx r / R$

To a first approximation, Weyl's analysis indicated a linear relation between redshift and distance for De Sitter's world.

The form of the relation between redshift and distance in De Sitter's universe thus obtained, was established using Weyl Principle.

On **May 23, 1923** Einstein sent a postcard to Weyl, and towards the end he wrote:[115]

"Regarding the cosmological problem I do not agree. According to De Sitter two points are accelerated to a sufficient distance away from each other [a motion of recession], if there is no quasi-static world, then away with the cosmological term".[116]

We may learn something of the reasoning behind the above quoted excerpt. The "cosmological problem" that Einstein did not agree with was probably *the De Sitter spectral shift effect.*

In 1917 Einstein explained that small velocities of the stars (with respect to the velocity of light), $v << c$, allow us to assume *a static universe.*[117] In Einstein's static universe no redshift of the spectral lines occurs unless the source moves with a velocity relative to the receiver. This result is consistent with the Doppler effect. If then a star moves through an otherwise static background of matter then it produces a redshift. Silpher's redshift data still suggested that $v << c$, allowing a quasi-static universe. In the De Sitter world, with the cosmological constant, we see a contribution to the redshift even if the emitting star does not move relative to the observer. It then looks as if the stars are accelerating away from the observer and are in a motion of recession. This result is the De Sitter spectral shift effect. In De Sitter's universe if all stars were supposed not to exist, with the exception of one single star placed at a huge distance from the observer, then the geometry of the world would have had effects on the signal sent from the source to the receiver. [118]

De Sitter's model was not without its difficulties; it had the strange De Sitter spectral shift effect. For Einstein its major weakness, to the point of apparent fatality, was simply that it violated his 1918 Mach's principle. If thus De Sitter's world was found to be non-static, then Einstein thought that there was no point in keeping the cosmological constant. After all Einstein introduced into his field equations the cosmological term having the cosmological constant as a coefficient, in order that the theory should yield a *static universe*.

### Friedmann's non-static solution 1922-1924

Aleksandr Friedmann published in the same year a model of an expanding universe and Einstein of course was not fond of it. On **May 29, 1922** Friedmann sent the article "On the Curvatures of Space" to *Zeitschrift für Physik* and it was received by the journal on **June 29, 1922**, and published on **September 13, 1922**.

Friedmann derived general models that could be obtained from Einstein's field equations and explained that: "from our hypothesis follows as special cases the cylindrical world of Einstein and the spherical world of de Sitter".[119]

According to Friedmann the spatial curvature of the universe is a function of time. If the curvature is independent of time then we get Einstein's and De Sitter's static models. We can assume a positive and negative value for the cosmological constant, and we can also consider the cosmological constant to be equal to zero. Friedmann thus discovered interesting non-static models with $\lambda = 0$ or $\lambda \neq 0$. This was a prediction of an expanding or a contracting universe. [120] Friedmann's model with $\lambda = 0$ was the simplest general relativity universe.[121] *Only in 1931 Einstein adopted this view and publically dropped the cosmological term.*

Later in a book in Russian Friedmann wrote about the results of the paper: "The stationary type of Universe comprises only two cases which were considered by Einstein and de Sitter. The variable type of Universe represents a great variety of cases".[122] Friedmann explained that cases of non-stationary universe were possible when the world's radius of curvature was constantly increasing in time, or else when the radius of curvature changes periodically; that is, Friedmann's solutions could be indeed either expanding or contracting universes.

There is the notable model which represents the expanding universe as a balloon (a closed universe). We take a deflated balloon and the fundamental particles of matter are represented by black dots distributed uniformly on this balloon. One particular dot may be marked so as to represent a given particle-observer. We blow up this balloon and as we do this we look at the movement of the black dots. They are all moving away from each other as we inflate the balloon; but the dots themselves are not actually moving, it is rather the material of the balloon that is expanding.

Einstein immediately replied to Friedmann's article by a note. His reply was received by *Zeitschrift für Physik* on **September 18, 1922**. Einstein's Bemerkung was published on **November 17, 1922**. He said that "the results concerning the non-stationary world contained in [Friedmann's] work, seem to me suspicious. In reality it turns out that the solution given in it does not satisfy the field equations". Einstein thought he found a mistake in Friedmann's results, which when corrected Friedmann's solution would give Einstein's static model.[123]

Of course Friedmann was disappointed when reading this note. On **December 6, 1922** Friedmann ventured to criticize Einstein's note, and he wrote a letter to Einstein: [124]

"Dear Professor,

From the letter of a friend of mine [Yuri A. Krutkov] who is now abroad [Krutkov came to Berlin on **September 27, 1922**] I had the honor to learn that you had submitted a short note to be printed in the 11[th] volume of the *Zeitschrift für Physik* [the noted was received on **September 28, 1922**], where it is stated that if one accepts the assumptions made in my article 'On the Curvature of Space', it will follow from the world equations derived by you that the radius of curvature of the world is a quantity independent of time [static world…]

Considering that the possible existence of a non-stationary world has a certain interest, I will allow myself to present to you here the calculations I have made".

Friedmann then asked Einstein:

"Should you find the calculations presented in my letter correct, please be so kind as to inform the editors of the *Zeitschrift für Physik* about it; perhaps in this case you will publish a correction to your statement or provide an opportunity for a portion of this letter to be published".

By the time this letter reached Berlin, Einstein had already left on a trip to Japan. In **April 1923** Einstein was invited to Leiden to attend the farewell lecture of Lorentz, who was about to retire. At the same time Krutkov was in Leiden, too – Einstein met with him at Ehrenfest's place, where he always stayed when coming to Leiden. Hence, only in the beginning of **May 1923** was he informed about the letter by Krutkov, and immediately was willing to write a second note to the *Zeitschrift für Physik*, received on **May 21, 1923** (published on **June 29, 1923**). [125]

Einstein was willing to correct the slip in his previous note: "In my previous note I have criticized the cited work [Friedmann's 1922 work, 'On the curvature of Space'], but my objection, as I became convinced by Friedmann's letter communicated to me by Mr. Krutkov, rested on an error in my calculations. I consider that Mr. Friedmann's results are correct and shed new light. It follows that the field equations, besides the static solution, permit dynamic (that is, varying with the time coordinate) spherically symmetric solutions for the spatial structure".[126]

In fact, Einstein was little impressed by Friedmann's mathematical models. In Einstein's draft of the second note to the *Zeitschrift für Physik*, in which he withdrew his earlier objection to Friedmann's dynamical solutions to the field equations, he crossed-out the final last section of the sentence, "a physical significance can hardly be ascribed to them", before sending the note to the editor of the *Zeitschrift für Physik*; thus Einstein originally wrote in the draft: "It follows that the field equations, besides the static solution, permit dynamic (that is, varying with the time coordinate) spherically symmetric solutions for the spatial structure, but a physical significance can hardly be ascribed to them".[127] It should be added that probably in conversations Einstein admitted the mathematical infallibility of Friedmann's calculations, and yet he added the same phrase.[128] We can thus almost hear Einstein thinking aloud: this work is just extra mathematical complication.

Friedmann stayed in Berlin in **August-September 1923**. Of course he attempted to meet with Einstein personally, but the meeting with Einstein never materialized (either during this visit to Germany or during the next one); Einstein was out of town on a vacation. On **September 13, 1923** Friedmann wrote a colleague: "Everybody was much impressed by my struggle with Einstein and my eventual victory, it is pleasant for me because of my papers; I shall be able to get them published more easily".[129]

Apparently neither the expanding universe, nor the contracting one, could be said to cause Einstein any interest in dynamical models in 1922-1923. The dynamical models seemed to be as remote as possible from Mach's ideas, and they rendered Einstein's static model no longer unique.[130] But it must be pointed out that in 1931, Einstein was going to realize that in the dynamical case (i.e. expanding universe) one could give up the cosmological constant; when dealing with a world which was not quasi-static, then Einstein, preferring a simple theory, thought that one did not need the cosmological term.

### Lamaître non-static solution 1925-1927

**In 1927** Lamaître had independently published quite the same model of the expanding universe as Friedmann. When Lemaître published his paper in Brussels in 1927, he was unaware of the little known papers of 1922 and 1924 by Friedmann in which he presented almost the same mathematical model as his.[131]

Lemaître became aware of Friedmann's work half-a-year after his own paper had appeared, when he met Einstein for the first time at the 1927 Solvay Conference in Brussels; there according to Lemaître Einstein told him: "Your calculations are correct, but your physics is abominable", and Einstein pointed out that his model had already been suggested by Alexander Friedman. Einstein's response to Lemaître's work indicated the same unwillingness to change his position that characterized his response to Friedmann's work. When Einstein met Lemaître he was willing to accept Lemaître's mathematics, but not the physics of the expanding universe.[132]

In a truly ironical manner William McCrea summarized Friedmann's and Lemaître's achievements: [133] "As a matter of history, therefore, the most extensive property of the universe ever discovered was successfully predicted by relativity theory". But Einstein was uneasy with the physics of Lemaître.

Before Lemaître's famous paper presenting his model of the expanding universe, one dealing with De Sitter's world was published **in 1925**. **In 1924-1925** Lemaître was a Ph.D. candidate at the Massachusetts Institute of Technology, doing work on problems in theoretical astrophysics and general relativity. Before traveling to the U.S, Lemaître spent the academic year 1913-1914 with Eddington at Cambridge Observatory, in Cambridge. [134]

In the 1925 paper Lemaître suggested a modification of De Sitter's model: [135]

"De Sitter's coordinates introduced a spurious inhomogeneity" a singularity at the equator in the coordinates used by De Sitter in 1917 for r = 0, and r = (π/2)R [136] "of the field which is not simply the mathematical appearance of center of an origin of coordinates, but really attributes distinct absolute properties to a center".

Lemaître "tried to remove the difficulty by introducing other coordinates and" was "led to a homogenous field; but, first, *the field is not static and, secondly, space has no curvature*". Namely, space is Euclidean.

Lemaître could probably accept the first point in De Sitter's universe. He quoted Eddington who wrote in his 1923 book on this subject: "It is sometimes urged against the de Sitter world that it becomes non-static as soon as any matter is inserted in it. But this property is perhaps in favor of the de Sitter theory rather than against it".[137]

Lemaître explained that the above treatment with new coordinates "evidences this non-static character of de Sitter's world which gives a possible interpretation of the mean receding motion of spiral nebulae". Lemaître's suggestion for a modification of De Sitter's model included a non-static character and dependence upon distance of the redshift of radiation caused by the Doppler effect.[138]

Nevertheless, Lemaître was unsatisfied with the suggested modification of De Sitter's model that he was advancing, because "we are led back to Euclidean space and to the impossibility of filling up an infinite space with matter which cannot but be finite". He thus concluded: "De Sitter's solution has to be abandoned, not because it is non-static, but because it does not give a finite space without introducing an impossible boundary".

In **the 1927** paper, "A homogeneous Universe of Constant Mass and Increasing Radius Accounting for the Radial Velocity of Extra-Galactic Nebulae", Lemaître compared actual measurements of galactic redshifts with predictions of a relativistic world model. Lemaître like Friedmann showed that the theory of relativity demanded that the universe in the large may be changing with time.

Although Lemaître's equations were the same as Friedmann's, Lemaître's work differed from the latter in some important respects.[139] The difference between the work of Friedmann and Lemaître lies more in their approach and spirit than in their formal content. Lemaître attempted to develop a physical cosmology and explained his models in terms of physical entities (stars, nebulae, etc) and connected his explanation with the redshifts of nebulae (galaxies). Lemaître described his model in terms of an expanding universe in which recession of the galaxies cause their received light to be redshifted. Friedmann proposed a general mathematical model to Einstein's field equations and cosmological constant (positive, negative, zero).[140]

Lemaître *demonstrated that the Einstein model is unstable so that when disturbed in the direction of expansion, it would go on expanding forever, and tending toward a De Sitter model in the limit.* Thus Lemaître made the best of both worlds, Einstein's and De Sitter's, by showing that they are initial and final states of a single more general model.[141] Lemaître explained this:[142]

"It seems desirable to find an intermediate solution which could combine the advantages of both [solutions: Einstein's and de Sitter's].

At first sight, such an intermediate solution does not appear to exist. A static gravitational field for a uniform distribution of matter without internal stress has only two solutions, that of Einstein and that of de Sitter. […] It is remarkable that the theory can provide no mean between these two extremes.

The solution of the paradox is that de Sitter's solution does not really meet all the requirements of the problem". Considering the De Sitter model that was first thought of as static model and the field was found to be no longer static, its radius no longer constant, but varying with time, "In order to find a solution combining the advantages of those of Einstein and de Sitter, we are led to consider an Einstein universe where the radius of space or of the universe is allowed to vary in an arbitrary way".

From 1922 to 1926 Edwin Hubble proposed a classification system for nebulae, both galactic and extragalactic. During 1926-1929 Hubble first verified that the galaxies are island universes and external to the Milky Way. Hubble then went to study the way galaxies were distributed in distance. If they increased in numbers in proportion to the survey volume, they would, then, clearly be the basic unit of the distribution of the universe. Hubble calculated the rate of increase in galaxy numbers with increasing volume.[143]

## New experimental findings

In **January 17, 1929** Edwin Hubble at Mount Wilson in California, announced the discovery that the actual universe was apparently expanding, as had been tentatively foreshadowed by the work of others, especially the work of Slipher. [144]

Hubble had observed that lines in the spectrum of each of nearly 40 "spiral nebulae" show redshift interpretable as the Doppler effect of a velocity of recession V (different in each case). Hubble observed more such objects. For each he estimated also the distance D. Hubble found that within the uncertainties of his determinations and within the resulting range of values of V and D these obeyed a simple linear relation: V = H·D, where H was the same for all the galaxies studied. We call this the *Hubble Law* and H the Hubble constant; the Hubble constant gives the rate of expansion. Hubble found H ≈ 500 kms-1Mpc-1.

Writing this equation in the equivalent form: D = TV, we call T the *Hubble time*. Hence V is the speed relative to the observer, i.e. the velocity of the galaxy concerned is essentially recession with speed V along the sight-line. Then it is evident that an observer on any one of all the galaxies to which Hubble's law applies must describe all the rest as obeying the same Hubble law relative to himself. Thus Hubble's achievement was the experimental discovery of the expansion of the universe.[145]

There was great excitement among physicists and astronomers when in 1929 Hubble announced the discovery that the actual universe is apparently expanding. Cosmologists thought that this seemed to be the greatest feat of theoretical prediction ever achieved.[146]

Around 1929 Cosmologists evidently accepted that general relativity could lead to two or three solutions: Einstein's static universe, De Sitter's stationary universe, and Minkowski's flat metric. That is why it was considered a tremendous achievement to: "[…] find that the only possible *stationary* cosmologies – i.e., the intrinsic properties of which are independent of time – are in fact those of Einstein and de Sitter, and that they arise from particular cases of a class of solutions whose general member defines a non-stationary cosmology".[147] Cosmologists felt that there was vast amount more that could be written on Einstein's and De Sitter's model's, on the question of finding the model for the universe, and studying these two possibilities; and so Richard Tolman remarked in the final line of his 1929 paper: "The investigation of non-static line-elements would be very interesting".[148]

Neither of the two stationary solutions, Einstein's and De Sitter's, proved to describe the experimental situation, because Einstein's static universe contained matter but no motion (it was truly static both in space and time). De Sitter's stationary universe did not because it had no matter, but as already stated curiously did have spectrum shifts (both red and blue) of test particles placed in the space which it described. This was due to a space-dependent factor in the metric coefficient of the time dimension, despite the so-called static nature of the space coordinates. Finally, "The De Sitter spectral shift effect" had been studied by Lamaître without convincing success. Hubble ended his pioneering paper, "A relation between distance and radial velocity among extra-galactic nebulae", by saying: [149]

"The outstanding feature, however, is the possibility that the velocity-distance relation may represent the de Sitter effect, and hence that numerical data may be introduced into discussions of the general curvature of space. In the de Sitter cosmology, displacements of the spectra arise from two sources, an apparent slowing down of atomic vibrations and a general tendency of material particles to scatter. The latter involves an acceleration and hence introduces the element of time. The relative importance of these two effects should determine the form of the relation between distances and observed velocities; and in this connection it may be emphasized that the linear relation found in the present discussion is a first approximation representing a restricted range in distance".

Remember that Hermann Weyl did pioneering work in the De Sitter spectral shift effect. He studied the De Sitter world of four-dimensional hyperboloid even after: "Recent observations made on spiral nebulae [by Hubble and published in 1929] have associated their extra-galactic nature and confirmed the redshift of their spectral lines as systematic and increasing with distance. By these facts the cosmological questions about the structure of the world as a whole, to which the general theory of relativity gave rise in purely speculative form, have acquired an augmented and empirical interested". [150]

Weyl does not mention Friedman in the 1923 fifth edition of *Raum, Zeit, Materie*, nor in his article "Redshift and Relativistic Cosmology" of 1930. And he does not mention Lemaître in his 1930 paper either. The first sign of Weyl noticing Friedmann's and Lemaître's work was in a lecture given in July 1933, where he adopted Eddington's point of view. He then thought that the true solution would lie somehow in between De Sitter's and Einstein's worlds, and he mentioned the solutions that had been given already in 1922 by Friedmann, and later by Lemaitre. Friedmann's solutions thus did not exist for Weyl until 1933. Einstein's world and De Sitter's universe were the only two exact solutions of cosmological siginificance for him. [151]

**The Lemaître-Eddington model**

About the **year 1930** Arthur Eddington studied non-static solutions, in particular, Lemaître's suggestion. McCrea wrote: "In Britain at any rate, part of the reason for the ready acceptance of the notion of the expansion of the universe was that it fitted in with ideas that A.S Eddington was developing at the time about the meaning of the constants of physics and about the harmonization of quantum physics and cosmic physics. In his scheme the cosmical constant $\Lambda$ played a crucial role, and he saw in the expansion an empirical means to evaluate this constant. Eddington had enormous prestige as a theorist, and Hubble had very great prestige as an observer. If they agreed the universe is expanding, then it had to be expanding". [152]

In 1930 Eddington criticized Einstein cosmological model, and found it was unstable; he searched for a new solution, but found that Lemaître had already solved the problem in 1927. Following Lemaître Eddington considered it essential to retain the cosmological constant and to treat the expanding universe as having started as an Einstein static universe. Eddington started from the (unstable) static Einstein model and expanded until it asymptotically approached the De Sitter model: [153]

"1. Working in Conjunction with Mr. G. C. McVittie, I began some months ago to examine whether Einstein's spherical universe is stable. Before our investigation was complete we learnt of a paper by Abbé G. Lemaître which gives a remarkably complete solution of the various questions connected with the Einstein and de Sitter cosmogonies. Although not expressly stated, it is at once apparent from his formulae that the Einstein world is unstable – an important fact which, I think, has not hitherto been appreciated in cosmogonical discussions. Astronomers are deeply interested in these recondite problems owing to their connection with the behavior of spiral nebulae; and I desire to review the situation from an astronomical standpoint, although my original hope of contributing some definitely new result has been forestalled by Lemaître's brilliant solution".

On a Friday **May 9, 1930** meeting of the Royal Astronomical Society the president called Prof. Eddington to give an account of his paper entitled "On the Instability of Einstein's Spherical World". Eddington gave the following report: [154]

"Some time ago I conjectured that Einstein's spherical world might be unstable. More recently I thought I saw a way to settle the question mathematically. I was working on this problem with Mr. McVittie, and we had nearly reached the solution when I learnt of a remarkable paper by Abbé G. Lemaître, of Louvain, published in 1927, which contained all the necessary mathematics. He does not say explicitly that Einstein's world is unstable, but it follows immediately from his equations. I think this makes a great difference to our outlook; Einstein's solution gave the only possible condition of equilibrium of the universe, and now this proves to be unstable. De Sitter's is also reckoned technically as an equilibrium solution, but it is a bit of a fraud; being entirely empty, there is nothing in his world whose equilibrium could possibly be upset. In saying this I am not disparaging it, because it is much more interesting than a genuine equilibrium solution would have been.

To discuss stability we must have a range of solutions, and Lemaître's work provides this. He treats of a world whose radius if a function of the time. Instead of having to choose between Einstein's and de Sitter's worlds our conclusion now is that the universe started as an Einstein world, being unstable it began to expand, and it is now progressing towards de Sitter's form as an ultimate limit".

Eddington, who was once the secretary of the *Monthly Notices of the Royal Astronomical Society*, arranged for a translation for Lemaître's French original 1927 paper from the *Annales de la Socie´te´ Scientifique de Bruxelles* into English for publication in the *Monthly Notices*. The English translation was published in **March 1931**. [155]

A few key paragraphs were deleted in the English translation, notably the ones dealing with Hubble's law, the velocity-distance relation and radial velocities and distances of nebulae, the estimations of the rate of expansion of the universe. It is significant to note that, in the paper of 1927 Lemaître does not mention Friedmann's work, but in the English translation of 1931 the "translator", or rather the corrector of the translation, gives items in a "References" list, one of which is Friedmann's 1922 paper. [156]

Recently it has been discovered by Mario Livio, who read the correspondence of Lemaître with the editor of *Monthly Notices* William Marshall Smart that, it was Lemaître himself who supplied the translation of his original paper to English, and he purposely omitted these paragraphs because of Hubble's 1929 work. [157]

On **February 17, 1931**, Smart wrote Lemaître: "Personally and also on behalf of the Society I hope that you will be able to do this" translation. On **March 9, 1931**, Lemaître replied: "I send you a translation of the paper […] I made this translation as exact as I can, but I would be very glad if some of yours would be kind enough to read it and correct my english which I am afraid is rather rough. No formula is changed, and even the final suggestion which is not confirmed by recent work of mine has not be modified". [158]

It appears that indeed Eddington *corrected* the translation. It seems that William McCrea, who knew personally several of the creators of modern astronomy and cosmology, especially the British members of the Royal Astronomical Society, confirms this:

"But their [Friedmann's and Lemaître's] work was scarcely noticed until Eddington noticed its significance. Later he caused a translation of Lemaître's paper (which acknowledged Friedmann's contribution) to be published in *Monthly Notices* **91**, 1931". [159]

Actually, Hubble (and Milton Humason) was very anxious to protect his priority in the discovery of the linear redshift-distance relation, but he did not engage in a debate on credit for the discovery of the expanding universe. [160]

In 1930 following Lemaître's lead Eddington suggested studying solutions which were not static: [161]

"2. *Expanding Universes*. – An infinite variety of solutions can be found representing spherical worlds which are not in equilibrium. Whilst remaining spherical they expand or contract, the radius (in terms of our ordinary standards which are in a constant, though unknown, relation to the cosmical standard $1/\lambda$) being a function of the time. In an expanding spherical world [dynamical model rather than static] the galaxies, since they continue to fill space uniformly, must become further apart as time progresses. Expanding solutions are therefore of astronomical interest as a possible explanation of the observed scattering apart of the spiral nebulae".

The Lemaître-Eddington model arises, as stated, from the perturbed Einstein static universe. For, as Eddington has pointed out, the Einstein model is unstable against a small change in the radius of curvature R. A decrease in R leads to contraction, whereas a slight increase of R leads to unlimited expansion. In the Lemaître-Eddington model the latter case is assumed. The rate of expansion is at first very slow, but then the model tends asymptotically to the De Sitter empty universe.

The Lemaître-Eddington model depended on the introduction of the cosmological constant which, Einstein was soon going to drop because of the latest experimental findings of Edwin Hubble. [162] Hence, Einstein objected to the Lemaître-Eddington model!

### Einstein drops the cosmological constant

In the years beyond 1930, the tide turned in favor of dynamical models of the universe. On **January 10, 1930** De Sitter appeared at the meeting of the Royal Astronomical Society, and later also in **summer 1930**, Hubble's 1929 paper led De Sitter to admit:[163]

"In *B.A.N.* [*Bulletin of the Astronomical Institutes of the Netherlands*] 185 it was pointed out that neither of the two possible static solutions of the differential equations can represent the observed facts of the finite density of matter in space and a systematic velocity of recession of the extragalactic nebulae proportional to the distance, and mention was made of the non-static solution found by Dr. Lemaître, which is compatible with these two observed facts. In the present article I will discuss some of the consequences of this solution, and will begin by recapitulating it in a notation slightly different from Lemaître's own". De Sitter has now vacillated towards Lemaître's solution.

In 1931 Einstein became aware of the experimental revolution during a visit to Caltech in Pasadena. On **January 29, 1931** he went to "The Monastery", Mount Wilson, to view the skies through the colossal telescope. After a twenty-mile drive to reach the observatory situated atop of the San Gabriel Mountains, Einstein arrived. Hubble and Dr. Walter S. Adams, director of the observatory accompanied Einstein to a fifty-foot-high movable observation platform beneath the renowned Hooker Telescope, at the time the world's largest, where Einstein viewed the heavenly bodies. Einstein was deeply impressed, and examined high-resolution spectrographs and other evidence that demonstrated that the universe was expanding. Hence, Einstein heard from Hubble about observational results; he commented that his "cosmological constant was superfluous".[164]

On **March, 1, 1931**, Einstein wrote Besso from Pasadena: [165]

"The people at Mount Wilson Observatory are excellent. They have recently found that the spiral nebulae are spatially approximately uniformly distributed and show a strong Doppler effect proportional to their distance, which follows without constraint from the theory of relativity (without cosmological constant)".

Einstein had occasion to read cautiously with Krutkov Friedmann's December 6, 1922 letter.[166] In that circumstance he was thinking just in terms of admitting the mathematical infallibility of Friedmann's calculations, but all that he could do about Friedmann's results is to say (and regret and delete before printing) that physical significance can hardly be ascribed to them.

Einstein had good Machian reasons for his reservations about dynamical solutions and in particular as regards Friedmann's model. Fairly obviously he read carefully Friedmann's paper showing that, we can assume a positive and negative value for the cosmological constant, and we can also consider the cosmological constant to be equal to zero, Friedmann's model with a cosmological constant $\lambda = 0$ is the simplest dynamical model of an expanding universe.

Aside for his Machian reservations, were it not for what Hubble had shown Einstein in Mount Wilson, Friedmann's model might still seem to him with no physical significance. But Hubble's discovery could be explained now without the cosmological constant and with the unmodified field equations, the simple and natural equations of the 1916 general relativity. It was very typical to Einstein that he used to do a theoretical work and he cared about experiments and observations.

At this point Einstein indeed thought that it ought to be possible to find dynamical solutions to his unmodified field equations. Means other than the cosmological constant had to be sought, namely, no means at all; and so he adopted Friedmann's model with a cosmological constant $\lambda = 0$, and searched for an expanding universe dynamical solution to his unmodified field equations.

### The Einstein-De Sitter universe

Upon his return to Berlin from Pasadena in 1931, Einstein, who was usually trusted as the authority of scientific matters, took the view that the most significant feature of the observed expansion of the universe for relativity was that it allowed him to discard the cosmological constant. Einstein had introduced it only so that the theory would admit a static universe; if the actual universe was now seen to be non-static, then the cosmological constant was simply unwanted. He therefore withdrew his support for the cosmological constant, and suggested that it should be dropped from the field equations of his theory. Einstein thus returned to the unmodified field equations of general relativity.[167]

Following Hubble's findings, of which Einstein supported, he immediately published a short paper in **April 1931**, "On the Cosmological problem of General Relativity". The paper began with Einstein's discussion of the problem shortly after the formation of general relativity in 1916, succeeded by numerous studies of opposite standpoint; and new experimental facts, with the results regarding Doppler shifts and the distribution of extra-galactic nebulae, making it clear that new ways of study in relativity were now opened. [168]

Einstein vacillated towards Friedmann's solution with a cosmological constant $\lambda = 0$;[169] and this amounted mainly to withdrawing the cosmological term, the "λ-Gliedes"[170]. Einstein then studied the following simple line element:[171]

$$(6)\ ds^2 = -R^2[dx_1^2 + \sin^2 x_1\, dx_2^2 + \sin^2 x_1\, dx_2^2 dx_3^2] + c^2 dx_4^2$$

R being a function of $x_4$ (or t) alone. Since this line element corresponded to $\lambda = 0$, it was just natural for Einstein to end the paper by saying that general relativity can justify the new findings of Hubble "without the λ-term".[172]

Einstein is reported to have said a year before his death: "Every man has his own cosmology and who can say that his own theory is right!"[173] And Einstein's cosmology was a cosmology "without the λ-term".

During the **autumn of 1931**, at a meeting of the British Association for the Advancement of Science, a gathering of the majority of practicing cosmologists certified Lemaître's model as the most likely correct description of the universe.[174] It culminated in the popular astronomical literature; popular papers flourished and the latest contributions were dedicated to Lemaître's model. In an article communicated by Eddington, Lemaître wrote: [175]

"Popular expositions [of his theory] have been given by:
G. Lemaître, 'La grandeur de l'espace,' *Revue des questions scientifiques*, March 1929.
W. De Sitter, 'The Expanding Universe,' *Scientia*, Jan, 1931."

In **January 1931** perhaps De Sitter supported the Lemaître-Eddington model with $\lambda > 0$ to which Einstein objected, but a few months later De Sitter read Einstein's 1931 paper and changed his mind. He adopted Einstein's new line element (1) and studied the non-static Einstein solution of the field equations with constant density. In a paper from **August 7, 1931**, De Sitter wrote:[176]

"The non-static solutions of the field equations of the general theory of relativity, of which the line element is,

$$(6a)\ ds^2 = -R^2 d\sigma^2 + c^2 t^2$$

R being a function of t alone, and $d\sigma^2$ being the line-element of a three-dimensional space of constant curvature with unit radius, have been investigated by Friedmann in 1922 and independently by Lemaître in 1927, and have attracted general attention during the last year or so. Einstein has lately [in his 1931 paper] expressed his preference for the particular solution of this kind corresponding to the value $\lambda = 0$ of the 'cosmological constant'. This solution belongs to a family of solutions which were not included in my discussion in B. A. N 193".[177]

In the 1930 paper (number 193) De Sitter indeed had not studied the non-static solution (1), but De Sitter's 1931 research led to the Einstein-De Sitter universe. A few months later, in 1932, in a joint paper, and in what seemed as a final end to the good old competing static models, Einstein and De Sitter presented the Einstein-De Sitter suggestion following Einstein's lead without the λ-term:[178]

"Historically the term containing the 'cosmological constant' λ was introduced into the field equations in order to enable us to account theoretically for the existence of a finite mean density in a static universe. It now appears that in the dynamical case this end can be reached without the introduction of λ."

De Sitter received a copy of Otto Heckmann's 1931 publication and then, when both he and Einstein were visiting Mount Wilson Observatory on **January 25, 1932**, they wrote the 1932 joint paper, and mentioned Heckmann:[179]

"In a recent note in the *Göttinger Nachrichten*, Dr. O. Heckmann has pointed out that the non-static solutions of the field equations of the general theory of relativity with constant density do not necessarily imply a positive curvature of three-dimensional space, but that this curvature may also be negative or zero.

There is no direct observational evidence for the curvature, the only directly observed data being the mean density and the expansion, which latter proves that the actual universe corresponds to the non-statical case. It is therefore clear that from the direct data of observation we can derive neither the sign nor the value of the curvature, and the question arises whether it is possible to represent the observed facts without introducing a curvature at all".

If the field equations were those given by Einstein's theory, the Friedman-Lemaître cosmological models resulted; the simplest of these was the Einstein-De Sitter model[180]: the model was the line-element (1) "without the introduction of λ" and "without introducing a curvature at all… If we suppose the curvature to be zero […] we neglect the pressure". Einstein and De Sitter derived the coefficient of expansion $h^2 = 1/3 \, \kappa\rho$, which "depends on the measured redshifts". And the numerical value of $\rho$ "happens to coincide exactly with the upper limit for the density adopted by one of us", by De Sitter in his August 1931 paper.[181]

## The Primeval Atom

At the time, on the other side of the Atlantic, Eddington was still intensely interested in the cosmological constant λ > 0. According to Eddington the effect of the cosmological constant was to introduce cosmical repulsion (provided λ > 0) into the universe, and this could be looked upon as the cause of the expansion of the universe.

The outcome of observations was therefore apparently most impressively to vindicate Einstein's introduction of the cosmological constant. Eddington's universe has no "beginning"; it arises, as already stated, from the perturbed Einstein static universe with no beginning; the Einstein static universe is unstable against a slight increase in the radius of curvature which leads to unlimited expansion. The rate of expansion is at first very slow, but then the model tends asymptotically to the De Sitter empty universe.

Eddington submitted the following paper (received on **August 11, 1931**), "On the Value of the Cosmical Constant" to *The Proceedings of the Royal Society of London*. He began his paper by saying that the cosmological term $\lambda$ "represents a scattering force which tends to make all very remote bodies recede from one another; this phenomenon is the basis of the theories of de Sitter and Lemaître concerning 'expansion of the universe'.

If the observed recession of the spiral nebulae is a manifestation of this effect the value of $\lambda$ can be found from the astronomical observations. Eddington found the value for the cosmological constant: "$\lambda = 9.79 \cdot 10^{-55}$, which gives a speed of recession of the spiral nebulae 528 km. per sec. per megaparasec. The observed speed according to Hubble is 465 km. per sec. per megaparsec."[182]

In **March 21, 1931** *Nature* published a paper by Eddington, "The End of the World: from the Standpoint of Mathematical Physics", in which he repeated his view that the universe had no beginning.

In **May 1931** Lemaître responded to Eddington's article and proposed the first modern history of the universe *having a clear beginning*. Lemaître suggested a model with point-source-creation in which the cosmological constant played an important role; the cosmological constant and the initial velocity of expansion were adjusted so that there was a stage in the expansion approximating to the Einstein static universe. Later stages in the expansion were then the same as in the Lemaître-Eddington model.[183]

However, because of Einstein's authority and since he withdrew the cosmological constant when dealing with dynamical models, the use of the cosmological constant was generally out of favor; hence, at the time Lemaître's ideas were not given much attention. By 1933 cosmologies have seen in his work as the beginning of big-bang cosmology. Taking a cosmological constant $\lambda > 0$, Lemaître found an expansion factor satisfying his equations for the relativistic homogeneous isotropic expanding cosmological model. The model had a singularity at time zero followed by rapid expansion, this being decelerated by self-gravitation leading to near-stagnation in the vicinity of the Einstein static state, independent of time, if the value of the cosmological constant $\lambda$ is suitably chosen, until the onset of accelerated expansion under cosmic repulsion. Lemaître pictured the very early universe as a "primeval atom", a cosmic atomic nucleus, with the big bang as its spontaneous radioactive decay. Thus the very early universe would have been dominated by high-energy particles producing a homogenous early universe. Cosmic rays were inferred to be the most energetic relict particles from the decay, so that they constituted background radiation for the model. Thus the very early universe would have been dominated by high-energy particles producing a homogeneous early universe. [184]

It was probably the second time that Lemaître's felt that his ideas had been rejected because of misunderstanding. André Deprit, a student of Lemaître, later mentioned: "For the past fifteen years, since his memorable address to the British Association in 1931, he had been put on the defensive. The Big Bang Theory had been held in suspicion by most astronomers, not the least by Einstein, if only for the reason that it was proposed by a Catholic priest and seconded by a devout Quaker, hence highly suspect of concordism".[185]

Indeed people had difficulties in accepting Big bang theory; in **January 1933**, Lemaitre and Einstein came to California for a series of seminars, there they met Hubble and others and discussed recent developments in cosmology. Journalists reported about the lectures, one of whom was a *New York Times* reporter who interviewed Lemaitre and wrote that, Lemaitre told his audience: "There is no conflict between religion and science".[186]

Although Eddington tried to push on in Lemaître's direction, in 1933 De Sitter was already confused, and perhaps following Einstein's lead he thought: "Astronomical observations give us no means whatsoever to decide which of these possible solutions corresponds to the actual universe. The choice must, as Sir Arthur Eddington says, depend on aesthetical considerations".[187] Einstein, however, had already made the "aesthetical" choice: he had dropped the cosmological constant.

On **January 23, 1932**, *The Brisbane Courier* reported:[188] "Sir Arthur [Eddington] went on to explain that light can no longer go round the world in finite time. The circumference of the world is expanding, he said, and light is like a runner on an expanding track with the winning post receding faster than he can run. In the early days light and other radiation went round and found the world until it was absorbed. The merry-go-round lasted during the very early stages of expansion. But when the world had expanded to 1.003 times its original radius the bell rang for the last lap. Light waves then running will make just one more circuit. Those which started later will never get round".

Eddington died in 1944 and left a finished draft of most of a book he had been writing, with brief notes of the intended contents of the rest. Eddington clearly meant this work to supersede most of his previous writings on the foundations of physics. The trouble was that nobody could understand it! Some people could follow what Eddington claimed to have done, and some could follow his mathematics, but none could see how everything fitted together. However, such was Eddington's prestige, as well as his success in deriving apparently from nowhere uncannily accurate values for all known fundamental constants of physics, that his colleagues thought he must surely be right, if only they could grasp his arguments. Hubble's own value of about 560 km s$^{-1}$ Mpc$^{-1}$ for the Hubble constant was available, and Eddington claimed to derive a theoretical value 572.36. In 1947 it was found that Eddington had made a mistake, he had overlooked a factor in his calculations that multiplied his value by 4/9.[189] After little more than a two decades of cosmological work Eddington's colleagues thought he must be surely right about the fundamental constants of nature; but what about another constant, the cosmological constant?

### Einstein's 70[th] Birthday and Mach's ideas

On Einstein's 70's Birthday Lemaitre contributed two articles.

In **July 1949** the *Reviews of Modern Physics* devoted an issue to a celebration of Einstein's 70's birthday. It contains articles about Einstein's achievements by Einstein's friends and colleagues. Lemaitre contributed a technical paper entitled, "Cosmological Application of Relativity", in which he explained in much detail the "Expanding Space-Friedmann's equation", and his 1931 new idea of cosmic rays. Lemaitre then mentioned in a short passage the central issue that long played an important role in Einstein's discussion with De Sitter: "It may be seen that de Sitter's singularity like Schwarzschild's singularity is an artificial singularity, not of the field but of the coordinates introduced to describe this field". This mention was made many years after the 1917 polemic with De Sitter; doubtless the elderly Einstein almost forgot about it (until Lemaitre perhaps reminded him some of the sweetest memories that were now perhaps too amusing to be forgotten).[190]

**In 1946** Paul Arthur Schilpp, a professor of philosophy, dedicated a book to Einstein that was later published in 1949, celebrating Einstein's 70's Birthday. Schilpp had edited a series of books about great living philosophers, and he wanted to edit a book on Einstein as well. Each book was devoted to a single man. It contained his specially written autobiography, followed by a series of essays by authorities evaluating and criticizing his work. These essays were then answered by the philosopher himself. Lemaître dedicated a popular essay to the cosmological constant.

With the aid of quotations from Eddington's 1933 book, *The Expanding Universe*,[191] and Einstein's works, Lemaître tried to persuade the reader that the structure of Einstein's field equations "quite naturally allows for the presence of a second constant besides the gravitational one", the cosmological constant. Lemaître explained: "The history of science provides many instances of discoveries which have been made for reasons which are no longer considered satisfactory. It may be that the discovery of the cosmological constant is such a case". The "reasons which are no longer considered satisfactory" were probably Einstein's static cosmological universe. Lemaître intended to demonstrate that "there are other empirical reasons to maintain lambda [the cosmological constant]", in a dynamical universe.[192]

Einstein did not agree that when dealing with an expanding universe, there were any reasons to maintain the cosmological constant. He explained that models of the expanding universe could be got without any mention of the cosmological constant, and he proposed that the cosmological term be again dropped from the theory of general relativity. Einstein replied to the article by Lemaître, and explained why he dropped the cosmological constant in the dynamical case:[193]

"I must admit that these arguments do not appear to me as sufficiently convincing in view of the present state of our knowledge.

The introduction of such a constant implies a considerable renunciation of the logical simplicity of theory, a renunciation which appeared to me unavoidable only so long as one had no reason to doubt the essentially static nature of space. After Hubble's discovery of the 'expansion' of the stellar system, and since Friedmann's discovery that the unsupplemented equations involve the possibility of the existence of an average (positive) density of matter in an expanding universe, that introduction of such a constant appears to me, from the theoretical standpoint, at present unjustified".

In 1946 (1949), it almost seemed pointless to Einstein to explain again the obvious that, the cosmological constant was superfluous. *Indeed Einstein's main object in the 1916 general theory of relativity was to develop a theory that the chosen path entered to it was psychologically the natural one, and its underlying assumptions would appear to have been secured experimentally*.

Expanding universe model could be achieved without the cosmological constant. However, Einstein still hoped that his theory would eliminate the epistemological weakness of Newtonian mechanics, the absolute space from physics. So far as "Mach's ideas", in the 1940s Einstein was not sure anymore that his theory eliminated the epistemological weakness of absolute space. The elderly Einstein could not say whether he was inspired by Mach as could the young Einstein who was inspired by Mach's ideas when creating the theory of relativity.

In 1948 Michele Besso (who recommended Mach to Einstein in 1897) asked Einstein about Mach's influence on his thought. Einstein replied to Besso from Princeton on **6 January 1948**: "As for Mach's influence on my own development, it has been very great. I certainly remember very well how, during my first years as a student, you directed my attention to his *Mechanik* and *Wärmelehre*, and both these two books made a deep impression on me. How far have they influenced my own work, frankly, it is not clear to me". Einstein was obviously not sure about this, and he ended the letter to Besso by saying, "As far as I can be aware, the indirect influence of D. Hume was greater on me [regarding special relativity]. I read this together with Conrad Habicht and Solovine in Bern. But as I said, I am not able to analyze the thinking rooted in the unconscious".[194]

Later this year Hermann Bondi and Thomas Gold came up with a new idea, namely, a steady-state theory of the expanding universe; there were two main elements to it, interwoven, a cosmological principle and a universe which is postulated to be homogeneous and stationary in its large scale appearance as well as in its physical laws. If we follow each element then we recognize the (dynamical) expanding universe and a version of Einstein's cosmological model. Over and above all, Bondi and Gold were fascinated by Mach's principle, and spoke about the difficulties "concerned with the absolute state of rotation of a body. Mach examined this problem very thoroughly and all the advances in theory which have been made have not weakened the force of his argument. According to 'Mach's Principle' inertia is an influence exerted by the aggregate of distant matter which determines the state of motion of the local frame of reference by means of which rotation of acceleration is measured".[195]

But the scientist who had most say in relativity matters in those days had already doubted Mach's principle in physics. Anyhow, "Mach's ideas" (economy of thought) seem to have still haunted Einstein **in 1950**:[196]

"What, then, impels us to devise theory after theory? Why do we devise theories at all? The answer to the latter question is simply: Because we enjoy 'comprehending,' i.e., reducing phenomena by the process of logic to something already known or (apparently) evident. New theories are first of all necessary when we encounter new facts which cannot be "explained" by existing theories. But this motivation for setting up new theories is, so to speak, trivial, imposed from without. There is another, more subtle motive. This is the striving toward unification and simplification of the premises of the theory as a whole (i.e., Mach's principle of economy, interpreted as a logical principle)".

Nevertheless, the most devout adherent of Mach's principle had to say in the year before his death: "In my opinion one should no longer speak at all of Mach's principle". And Einstein explained further that Mach's principle dated back to days when it was thought that matter was the only physical entity. He said that "I am well aware of the fact that I have been also influenced by this obsession for a long time".[197]

**Notes and References**

[116] Abraham Pais translated the last sentence: "If there is no quasi-static world, then away with the cosmological term". Pais, 1982, p. 288.

[117] Einstein, 1917, p. 143.
[118] Neuenschwander, 2008, p. 26.

[119] Friedmann, Alexander "Über die Krümmung des Raumes", *Zeitschrift für Physik* 10, 1922, pp. 377-386; p. p.378.
[120] McCrea, 1988, p. 53.

[121] McCrea, 1988, p. 53.

[122] Carmeli, Moshe (ed), *Relativity: Modern Large-scale Spacetime Structure of the Cosmos*, Singapore: World Scientific Publishing, 2008; p. 321.
[123] Einstein, "Bemerkung zu der Arbeit von A. Friedmann, 'Über die Krümmung des Raumes'," *Zeitschrift für Physik* 11, 1922, p. 326.
[124] Frenkel, Viktor, "Einstein and Friedmann", in Balashov, Vizgin, Vladimir,and Balashov, Yuri, *Einstein's Studies in Russia, Einstein Studies* Vol. 10, 2002, Boston: Birhäuser, pp. 1-15; p. 3; Carmeli, 2008, p. 321; Tropp, Eduard A., Frenkel, Viktor Ya. and Chernin, Artur D., Alexander A *Friedmann: The Man who Made the Universe Expand*, Cambridge: Cambridge University Press, 1993, p. 170.
[125] Einstein, "Notiz zu der Arbeit von A. Friedmann 'Über die Krümmung des Raumes'", *Zeitschrift für Physik* 16, 1923; Tropp, Frenkel, and Chernin, 1993, p. 171.
[126] Einstein, 1923, p. 228.
[127] Stachel, John, "Eddington and Einstein", *The Prism of Science. The Israel Colloquium: Studies in History and Philosophy, and Sociology of Science* 2, 1986, pp. 225-250, reprinted in Stachel, John, *Einstein from 'B' to 'Z'*, 2002, Washington D.C.: Birkhäuser, pp. 453-475; p. 469.
[128] Frenkel, 2002, p. 7.
[129] Quoted in Tropp, Frenkel, and Chernin, 1993, p. 174.
[130] McCrea, William, Hunter, "The Cosmical Constant", *Quarterly Journal of the Royal Astronomical Society* 12, 1971, pp.140-153; p.141.

[131] Friedmann, Alexander, "Über die Möglichkeit einer Welt mit konstanter negativer Krümmung des Raumes", *Zeitschrift für Physik* 21, 1924,  pp. 326-332.
According to Lemaître Einstein told him: "Vos calculs sont corrects, mais votre physique est abominable". Lemaître, Georges, "Rencontres avec A. Einstein", *Revue des questions scientifiques* 129, 1958, pp. 129-132; p. 129. Kragh, H. & Smith, R. W., "Who Discovered the Expanding Universe?", *History of Science* 41, 2003, p.141-162; p. 147.
[132] Kragh, Helge, *Cosmology and Controversy*, Princeton: Princeton University Press, 1996, p. 29-30.
[133] McCrea, William, Hunter, "Cosmology, A Brief Review", *Quarterly Journal of the Royal Astronomical Society*, Vol. 4, 1963, pp. 185-202; p. 186.
[134] Realdi, Matteo, *Cosmology at the turning point of relativity revolution. The debates during the 1920's on the "de Sitter Effect"*, Doctoral thesis, 2008, p. 113; Realdi, Matteo, Peruzzi, Guilio, "Einstein, de Sitter and the beginning of relativistic cosmology in 1917", *General Relativity and Gravitation* 41, 2009, pp. 225-247.
[135] Lemaître, Georges, "Note on de Sitter's Universe, *Journal of Mathematical and Physics* 4, May 1925, pp 188-192; p. 192.
[136] De Sitter, 1917a.

[137] Eddington, 1923, p. 161.

[138] Kragh, Helge, "Georges Lemaître", in Gillispie, Charles, *Dictionary of Scientific Biography*, 1970, New York: Scribner & American Council of Learned Societies, pp. 542–543.
[139] McCrea, 1971, pp.141-142.
[140] Lemaître, Abbé, G., "Un Univers homogène de masse constante et de rayon croissant rendant compte de la vitesse radiale des nébuleuses extra-galactiques", *Annales de la Societe Scientifique de Bruxelles* A47, 1927, p. 49-59; translation (adaptation): Lemaître, Abbé, Georges (1931a), "Expansion of the universe, A homogeneous universe of constant mass and increasing radius accounting for the radial velocity of extra-galactic nebulae", *Monthly Notices of the Royal Astronomical Society* 91, 1931, p.483-490; Kragh, 1996, p. 29-30.